\documentclass{emulateapj}

\usepackage{float}
\usepackage{amsmath}
\usepackage{epsfig,floatflt}
\usepackage{subfigure}

\begin{document}

\title{Power spectrum estimation from high-resolution maps by Gibbs
  sampling}

\author{H.\ K.\ Eriksen\altaffilmark{1}, I.\ J.\
  O'Dwyer\altaffilmark{2}, J.\ B.\ Jewell\altaffilmark{3}, B.\ D.\
  Wandelt\altaffilmark{4}, D.\ L.\ Larson\altaffilmark{5}, K.\ M.\ G\'orski\altaffilmark{6},
  S. Levin\altaffilmark{7}, A.\ J.\ Banday\altaffilmark{8} and P.\ B.\ Lilje\altaffilmark{9}} 

\altaffiltext{1}{Institute of Theoretical Astrophysics, University of
Oslo, P.O.\ Box 1029 Blindern, N-0315 Oslo, Norway; Centre of
Mathematics for Applications, University of Oslo, P.O.\ Box 1053
Blindern, N-0316 Oslo; Jet Propulsion Laboratory, M/S 169-327, 4800 Oak Grove Drive,
Pasadena CA 91109; California Institute of Technology, Pasadena, CA
91125; email: h.k.k.eriksen@astro.uio.no}

\altaffiltext{2}{Astronomy Department, University of Illinois at
  Urbana-Champaign, IL 61801-3080; email: iodwyer@astro.uiuc.edu}

\altaffiltext{3}{Jet Propulsion Laboratory, M/S 126-347, 4800 Oak
Grove Drive, Pasadena CA 91109; email: Jeffrey.B.Jewell@jpl.nasa.gov}

\altaffiltext{4}{Astronomy Department, University of Illinois at
  Urbana-Champaign, IL 61801-3080; Department of Physics, University
  of Illinois at Urbana-Champaign, IL 61801-3080; email:
  bwandelt@uiuc.edu}

\altaffiltext{5}{Astronomy Department, University of Illinois at
  Urbana-Champaign, IL 61801-3080; email: dlarson1@uiuc.edu}

\altaffiltext{6}{Jet Propulsion Laboratory, M/S 169-327, 4800 Oak
Grove Drive, Pasadena CA 91109; Warsaw University Observatory, Aleje
Ujazdowskie 4, 00-478 Warszawa, Poland; email:
Krzysztof.M.Gorski@jpl.nasa.gov}

\altaffiltext{7}{Jet Propulsion Laboratory, M/S 169-506, 4800 Oak
Grove Drive, Pasadena CA 91109; email: Steven.Levin@jpl.nasa.gov}

\altaffiltext{8}{Max-Planck-Institut f\"ur Astrophysik,
Karl-Schwarzschild-Str.\ 1, Postfach 1317, D-85741 Garching bei
M\"unchen, Germany; email: banday@MPA-Garching.MPG.DE}

\altaffiltext{9}{Institute of Theoretical Astrophysics, University of
Oslo, P.O.\ Box 1029 Blindern, N-0315 Oslo, Norway; Centre of
Mathematics for Applications, University of Oslo, P.O.\ Box 1053
Blindern, N-0316 Oslo; email: per.lilje@astro.uio.no}


\begin{abstract}
We revisit a recently introduced power spectrum estimation technique
based on Gibbs sampling, with the goal of applying it to the
high-resolution \emph{WMAP} data. In order to facilitate this
analysis, a number of sophistications have to be introduced, each of
which is discussed in detail. We have implemented two independent
versions of the algorithm to cross-check the computer codes, and to
verify that a particular solution to any given problem does not affect
the scientific results. We then apply these programs to simulated data
with known properties at intermediate ($N_{\textrm{side}} = 128$) and
high ($N_{\textrm{side}} = 512$) resolutions, to study effects such as
incomplete sky coverage and white vs.\ correlated noise. From these
simulations we also establish the Markov chain correlation length as a
function of signal-to-noise ratio, and give a few comments on the
properties of the correlation matrices involved. Parallelization issues
are also discussed, with emphasis on real-world limitations imposed by
current super-computer facilities. The scientific results from the
analysis of the first-year \emph{WMAP} data are presented in a
companion letter.
\end{abstract}



\section{Introduction}

The subject of cosmic microwave background (CMB) power spectrum
estimation has been a very active research field for many years now,
and the combined effort from the scientific community has resulted in
a number of qualitatively different methods. Broadly, one may
classify these methods into three groups, namely maximum likelihood
methods
\citep{gorski:1994,gorski:1996,tegmark:1997,bond:1998,oh:1999,dore:2001},
pseudo-$C_{\ell}$ methods \citep{wandelt:2001, hivon:2002,
hansen:2002,hansen:2003}, and specialized methods \citep{van
leeuwen:2002, challinor:2002,wandelt:2003}. In general the maximum
likelihood methods are more accurate than the (often Monte Carlo
based) pseudo-$C_{\ell}$ methods, but they are usually so at a
prohibitive computational cost. And even these methods can only return
very approximate summaries of the error bars, since exploring the
likelihood away from the peak is practically impossible. Further, the
specialized methods are usually only applicable following rather
restrictive assumptions. For general experiments we have up to now
been left with the rather uncomfortable choice between the optimal but
prohibitively expensive, and the feasible but approximate.

In this context, a method based on Monte Carlo Markov Chains and Gibbs
sampling was very recently developed by Jewell, Levin \& Anderson
(2004) and Wandelt, Larson \& Lakshminarayanan (2004) which may change
this picture. The fundamental idea behind this method is to solve the
power spectrum estimation problem by establishing its posterior
probability distribution through sampling, rather than by direct
solution of the corresponding optimization problem. In its most
general form, the scaling of this Monte Carlo method equals that of
the map making process, which is to be compared to the typical
$\mathcal{O}(N_{\textrm{pix}}^3)$ scaling for traditional maximum
likelihood estimators \citep{borrill:1997}, $N_{\textrm{pix}}$ being
the number of pixels in the map. Further, for an experiment with
spherically symmetric beams and uncorrelated noise, one may work
directly with maps instead of time-ordered data, in which case the
scaling reduces to that of a spherical harmonics transform, namely
$\mathcal{O}(N_{\textrm{pix}}^{3/2})$, using the
HEALPix\footnote{http://www.eso.org/science/healpix/} pixelization.

Until now, the only application of this method to cosmological data
was the analysis of the low-resolution \emph{COBE}-DMR data, presented
by \citet{wandelt:2004}. In the following we demonstrate the
practicality of this method for current and future experiments, as we
for the first time apply it to a large data set, namely the
\emph{Wilkinson Microwave Anisotropy Probe} (\emph{WMAP}) data
\citep{bennett:2003a}. This data set consists of eight cosmologically
important frequency bands, each with about three million pixels, and
it is therefore an excellent test bed for any new algorithm. We have
developed two independent implementations of the algorithm, one called
\emph{Commander}\footnote{Implemented by H. K. Eriksen and J. B.
Jewell.}  (``Commander is an Optimal Monte-carlo Markov chAiN Driven
EstimatoR'') and the other called \emph{MAGIC}\footnote{Implemented by
I. J. O'Dwyer, D. L. Larson and B.\ D.\ Wandelt.} (``Magic Allows
Global Inference of Covariance'') \citep{wandelt:2003b}. We have
tested these implementations extensively, and found that they produce
statistically identical results.

The goals of the present paper are twofold. First, we prepare for the
actual \emph{WMAP} analysis by developing a number of sophistications
to the Gibbs sampling algorithms necessary to facilitate a
high-resolution analysis. Second, we apply the computer codes to
simulated data in order to verify that the codes perform as expected,
and to build up an intuitive understanding of issues such as the
Markov chain correlation length vs.\ the signal-to-noise ratio,
multipole coupling vs.\ sky coverage, and sufficient sampling vs.\
overall CPU time. A proper understanding of these questions is crucial
in order to optimize real-world analyses. The scientific results from
the \emph{WMAP} analysis are reported in a companion letter by
\citet{odwyer:2004}.

\section{Algorithms}

This paper is a natural extension of the work presented by
\citet{jewell:2004} and \citet{wandelt:2004}, and we will in the
following frequently refer to those papers. Further, we do not attempt
to re-establish the motivation behind the Gibbs sampling approach
here, but refer the interested reader to those papers for details and
proofs. In the present paper we simply summarize the operational steps
of the algorithm, and specialize the discussion to the problems
encountered when analyzing the first-year \emph{WMAP} data.

We now define some notation. The data are given in the form of $N$ sky
maps (also called ``bands'' or ``channels'')
\begin{equation}
\mathbf{d}_k = \mathbf{A}_k \,\mathbf{s} + \mathbf{n}_k,
\label{eq:data_def}
\end{equation}
where $k=1,\ldots,N$ runs over the bands. $\mathbf{d}_k$ is the vector
of observed pixel values on the sky, $\mathbf{A}_k$ is the matrix
corresponding to beam convolution, $\mathbf{s}$ is the true sky
vector, and $\mathbf{n}_k$ is instrumental noise.

As mentioned above, our main scientific goal of the current work is to
analyze the first-year \emph{WMAP} data, and for that reason we assume
the beam to be azimuthally symmetric \citep{page:2003}. The beam
convolution $\mathbf{A}$ may therefore be computed in harmonic space
by a straightforward multiplication of the corresponding Legendre
components $b_{\ell}^k$. In order to simplify the notation, we
incorporate the pixel window function into $b_{\ell}^k$.

Further, for the \emph{WMAP} data, it is reasonable to approximate the
noise as uncorrelated, but non-uniform, so that the real space noise
covariance matrix can be written as $\mathbf{N}_{ij,k} =
\sigma^2_{i,k} \delta_{ij}$, where $\sigma_{i,k}$ is the noise
standard deviation of the $i$th pixel of the $k$th sky map. In fact,
we explicitly demonstrate the validity of this assumption in section
\ref{sec:highres_sim}, by first analyzing simulations including white
noise and then correlated noise, showing that they are statistically
consistent for the levels of correlated noise present in the
\emph{WMAP} data. Finally, we assume the CMB fluctuations to be
Gaussian and isotropic, and the signal covariance matrix therefore
simplifies considerably, $\mathbf{C}_{\ell m,\ell'm'} = C_{\ell} \,
\delta_{\ell\ell'} \, \delta_{mm'}$.

\subsection{Basic Gibbs sampling}
\label{sec:basic_gibbs}

The idea behind the Gibbs sampling power spectrum estimation technique
is to draw samples from the probability density
$P(C_{\ell}|\mathbf{d})$. The properties of this density can then be
summarized in terms of any preferred statistic, such as its
multivariate mean or mode. However, one of the major strengths of the
Gibbs sampling approach is that it allows for a global, optimal
analysis, and it should therefore not be considered as yet another
maximum likelihood technique, although it certainly is able to produce
such an estimate.

While direct sampling from the probability density
$P(C_{\ell}|\mathbf{d})$ is difficult, it is in fact possible to
sample from the joint density $P(C_{\ell}, \mathbf{s}|\mathbf{d})$,
and then marginalize over the signal $\mathbf{s}$. This is feasible
because the theory of Gibbs sampling tells us that if it is possible
to sample from the \emph{conditional} densities
$P(\mathbf{s}|C_{\ell}, \mathbf{d})$ and
$P(C_{\ell}|\mathbf{s},\mathbf{d})$, then the two following equations
will, after an initial burn-in period, converge to being samples from
the joint density $P(C_{\ell}, \mathbf{s}|\mathbf{d})$:
\begin{align}
\mathbf{s}^{i+1} &\leftarrow P(\mathbf{s}|C_{\ell}^{i}, \mathbf{d}), \\
C_{\ell}^{i+1} &\leftarrow P(C_{\ell}|\mathbf{s}^{i+1}).
\end{align}
Thus, given some initial power spectrum and the data, we may iterate
these two relations, discard the first few pre-convergence samples (if
necessary), and then use the remaining samples to construct whatever
statistic we prefer for the power spectrum. One further advantage of
this approach is that we probe the joint distribution, and we may
therefore quantify joint uncertainties. And, in the process, we also
obtain a Wiener filtered map which may be useful for other studies.

\begin{figure*}

\epsscale{0.7}
\plotone{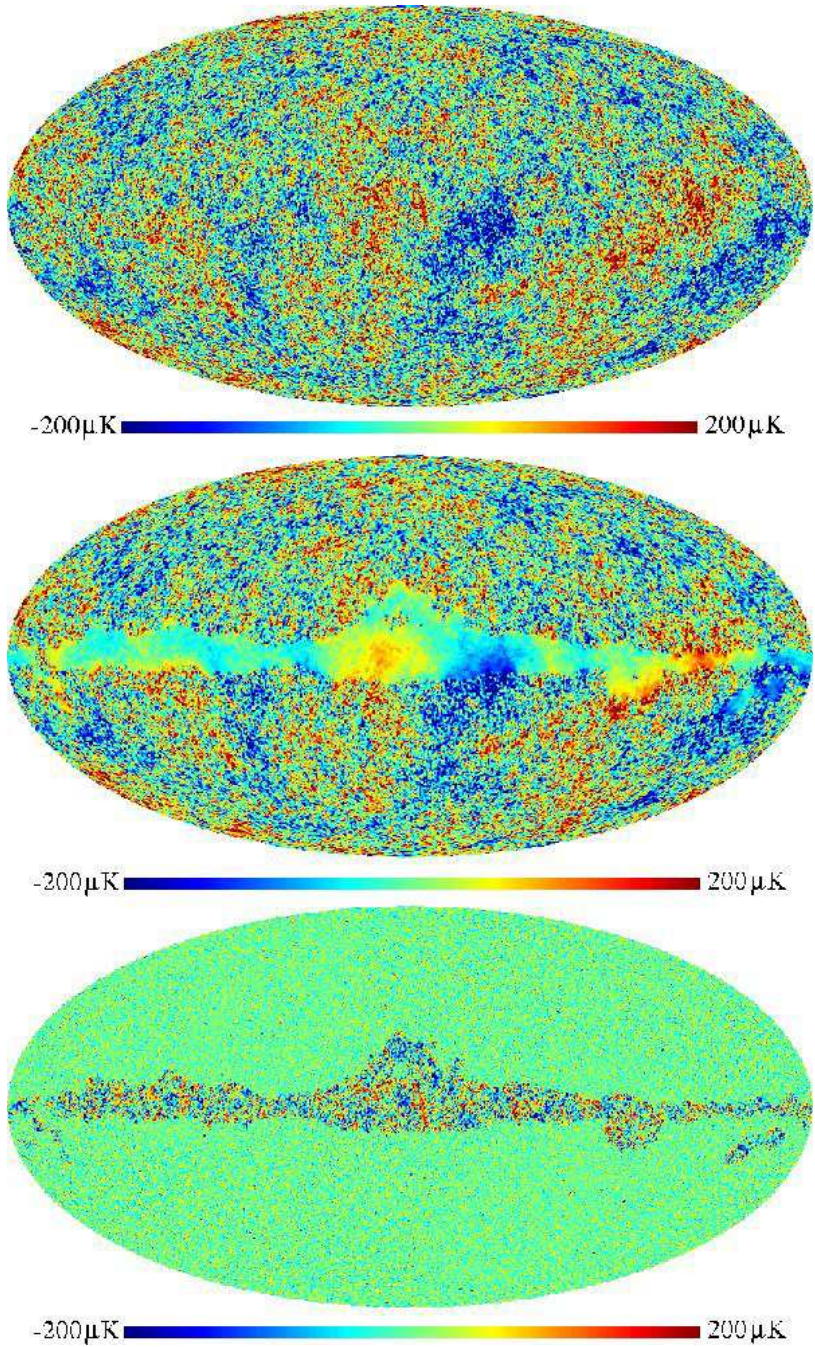}

\caption{Examples of the maps produced in one step of the Gibbs
 sampler. Top panel: The full-sky, noise-less Gibbs sample, $s$.
 Middle panel: The mean field (Wiener-filtered) map, $x$. Bottom
 panel: The fluctuation map, $y$.}
\label{fig:maps}
\end{figure*}

Drawing a power spectrum $C_{\ell}^{i+1}$ given a sky map
$\mathbf{s}^{i}$ is trivial (see, e.g., Wandelt et al.\ 2004). Given
the power spectrum of the signal map (often written on the form
$\sigma_{\ell} = \sum_{m=-\ell}^{\ell} |a_{\ell m}|^2$), one draws
$2\ell-1$ Gaussian random variates $\rho_{\ell}^j$ with zero mean and
unit variance, and form the sum $\rho_{\ell}^2 = \sum_{j=1}^{2\ell-1}
|\rho_{\ell}^j|^2$. The desired power spectrum sample is then given by
\begin{equation}
C_{\ell}^{i+1} = \frac{\sigma_{\ell}}{\rho_{\ell}^2}
\end{equation}

On the other hand, drawing a sky map $\mathbf{s}^{i}$ given the data
and an assumed power spectrum, is certainly not trivial. Again, we
refer the interested reader to the above-mentioned papers for
justification of the following procedure, and here we only review the
operational steps.

\begin{figure}
\mbox{\epsfig{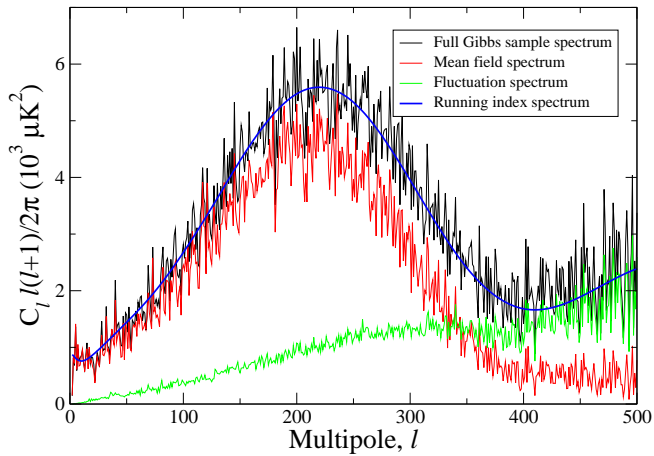}} 

\caption{Spectra corresponding to the maps in figure \ref{fig:maps}.
The red lines shows the spectrum of the Wiener-filtered map, and is a
biased estimate of the underlying spectrum. Therefore, the Gibbs
sampler adds a fluctuation term to the Wiener-filtered map, to yield
an unbiased estimate of the true spectrum.}
\label{fig:ex_spectra}
\end{figure}

The map sampling process is performed in two steps, the first being to
solve the following equation for the so-called mean field map
$\mathbf{x}$,
\begin{equation}
\begin{split}
\left(\mathbf{C}^{-1} + \left[\sum_{k=1}^{N}
  \mathbf{A}_k^{\textrm{T}} \mathbf{N}^{-1}_k
  \mathbf{A}_k\right]\right)\,\mathbf{x}
  = \sum_{k=1}^{N} \mathbf{A}_k^{\textrm{T}}
  \mathbf{N}^{-1}_k \mathbf{r}_k^{\textrm{s}}.
\end{split}
\label{eq:meanfield}
\end{equation}
Here $\mathbf{r}_k^{\textrm{s}} = \mathbf{d}_k$ is the residual signal
map. The reason for introducing this notation will become clearer when
additional components are introduced into the sampling chain. Any new
component we may wish to include in the analysis will simply be
subtracted from the data, to form an actual residual map from which
the mean field map is computed.

The mean field map is a generalized Wiener filtered map, and as such
is biased. To construct an unbiased sample one must therefore add a
fluctuation map $\mathbf{y}$ with properties such that the sum of the
two fields is a sample from the distribution of the correct mean and
covariance. The appropriate equation for this fluctuation map is
\begin{equation}
\begin{split}
\left(\mathbf{C}^{-1} + \left[\sum_{k=1}^{N}
  \mathbf{A}_k^{\textrm{T}} \mathbf{N}^{-1}_k
  \mathbf{A}_k\right]\right)\,\mathbf{y}
  = \quad\quad\quad\quad\quad\quad \\ \quad\quad\quad\quad\quad\quad
=\mathbf{C}^{-1/2}\mathbf{\omega}_0 + \sum_{k=1}^{N}
  \mathbf{A}_k^{\textrm{T}} \mathbf{N}^{-1/2}_k \mathbf{\omega}_k,
\end{split}
\label{eq:fluctmap}
\end{equation}
where $\mathbf{\omega}_k$ are Gaussian white noise maps of zero mean
and unit variance. 

Examples of such maps are shown in figure \ref{fig:maps}, and the
corresponding power spectra are shown in figure
\ref{fig:ex_spectra}. However, in practice the two equations are
solved simultaneously by solving for the sum of $\mathbf{x}$ and
$\mathbf{y}$, in order to reduce the total CPU time.

Finally, we point out that even though the Gibbs sampling technique is
a Bayesian method, a frequentist view may be taken by choosing a
uniform prior. In that case, the procedure reduces to simply exploring
the joint likelihood, and frequentist concepts such as the maximum
likelihood estimate may be established.

\subsection{Consistent treatment of mono- and dipole contributions}

One of the most elegant features of this formalism is its ability to
incorporate virtually any real-world complication, as discussed by
\citet{jewell:2004} and \citet{wandelt:2004}. A few examples of this
flexibility are applications to $1/f$ noise, asymmetric beams,
non-cosmological foregrounds, or arbitrary sky coverage. However, in
this paper we include only the effects of the mono- and dipole
contributions (which may be thought of as foregrounds) and that of
partial sky coverage, given that our main scientific goal is to
analyze the fairly well-behaved \emph{WMAP} data.

The question regarding mono- and dipole contributions has gained
renewed importance during the previous year, given the very active
debate concerning the quadrupole seen in the \emph{WMAP} data. This
quadrupole appears to be small compared to the best-fit cosmological
model \citep{spergel:2003,efstathiou:2003a,costa:2004}, and several
authors have considered what this may imply in terms of new
physics. However, the exact significance of this anomaly is difficult
to assess for several reasons, but mainly because of uncertainties in
the foreground subtraction process
\citep{eriksen:2004,slosar:2004}. Methodology issues for estimating
the lowest multipole amplitudes have also been pointed out
\citep{efstathiou:2003b}. Strongly related to both these issues is the
fact that non-cosmological mono- and dipole contributions may couple
into the other low-order modes through incomplete sky coverage.

The most common way of handling this latter problem is to fit a mono-
and dipole to the incomplete sky, including internal coupling caused
by the sky cut, and then simply subtract the resulting best-fit
components from the data. However, this procedure neglects the noise
correlations that are introduced by removing any fitted templates. The
Gibbs sampling framework allows a statistically more consistent
approach: rather than directly subtracting the fitted mono- and
dipoles from the data, one may marginalize over them through sampling,
and thus recognize the inherent uncertainties involved.

As always in Bayesian analyses, one has to choose a prior, and the
most natural choice in this case is a uniform prior. This corresponds
to saying that we do not know anything about these components. For
analytic computations and proofs, however, it is more convenient to
define this as a Gaussian with infinite variance, which is just a
different way of parameterizing a uniform prior. It should be noted
that a uniform prior does not mean that these components are
unrestricted, but, rather, it simply means that their values are
determined by the data alone.

Again, general formalisms for handling this type of problem were
described by \citet{jewell:2004} and \citet{wandelt:2004}, and we will
only repeat the operational steps here, in a notation suitable for our
purposes. Let us first define a $N_{\textrm{pix}} \times 4$ template
matrix $\mathbf{T}$ containing the four real spherical harmonics in
pixel space,
\begin{equation}
\mathbf{T} = \left(\mathbf{Y}_{00}, \mathbf{Y}_{1-1}, \mathbf{Y}_{10}, \mathbf{Y}_{11}\right),
\end{equation}
where $\mathbf{Y}_{\ell m} = \left(Y_{\ell m}(\theta_1,\phi_1), \ldots,
  Y_{\ell m}(\theta_{N_{\textrm{pix}}},
  \phi_{N_{\textrm{pix}}})\right)^{\textrm{T}}$ and
\begin{align}
Y_{00}(\theta,\phi) &= 1/\sqrt{4\pi} \\
Y_{1-1}(\theta,\phi) &= \sqrt{3/4\pi} \sin{\theta} \sin{\phi} \\
Y_{10}(\theta,\phi) &= \sqrt{3/4\pi} \cos{\theta}  \\
Y_{11}(\theta,\phi) &= \sqrt{3/4\pi} \sin{\theta} \cos{\phi}
\end{align}
Note that $\mathbf{T}$ is a projection matrix onto the subspace
spanned by the corresponding templates.

Next we define a vector of template amplitudes $\mathbf{w}_k =
(a^{k}_{00}, a^{k}_{1-1}, a^{k}_{10}, a^{k}_{11})^{\textrm{T}}$,
letting the amplitudes be different for each channel, since we have no
reason to assume that these components are frequency
independent. Thus, the mono- and dipole contribution to the $k$th
channel is $\mathbf{t}_k = \mathbf{T} \mathbf{w}_k$.

We now want to sample from the conditional distribution
$P(\mathbf{w}_k|\mathbf{d}_k,\mathbf{s})$, and this is done (assuming
the infinite variance prior) by solving the following equation,
\begin{equation}
\left(\mathbf{T}^{\textrm{T}} \mathbf{N}^{-1}_{k} \mathbf{T}\right)
\mathbf{w}_k = \mathbf{T}^{\textrm{T}} \mathbf{N}^{-1}_{k}
\mathbf{r}_k^{\textrm{md}} + \mathbf{\delta}_k,
\label{eq:mono_dipole}
\end{equation}
where the mono- and dipole residual map is $\mathbf{r}_k^{\textrm{md}}
= \mathbf{d}_k - \mathbf{A}_k \mathbf{s}$, and
\begin{equation}
\delta_k = 
\left[
\begin{array}{r}
\mathbf{Y}_{00}^{\textrm{T}}\;\;\; \mathbf{N}^{-1/2}_{k}\; \omega_k^{(1)} \\
\mathbf{Y}_{1-1}^{\textrm{T}}\, \mathbf{N}^{-1/2}_{k} \;\omega_k^{(2)} \\ 
\mathbf{Y}_{10}^{\textrm{T}}\;\;\; \mathbf{N}^{-1/2}_{k}\; \omega_k^{(3)} \\ 
\mathbf{Y}_{11}^{\textrm{T}}\;\;\; \mathbf{N}^{-1/2}_{k} \;\omega_k^{(4)}
\end{array}
\right].
\end{equation}
Here, $\mathbf{\omega}_k^{(i)}$ are white noise maps of vanishing mean and
unit variance.

The next step in traditional Gibbs sampling would now be to sample
from the conditional density
$P(\mathbf{p}_{\textrm{md}}|\mathbf{w}_k)$, where
$\mathbf{p}_{\textrm{md}}$ are the parameters of the probability
distribution describing the mono- and dipoles. However, since we have
chosen a very special prior, namely one with infinite variance, this
distribution does not change, and no sampling is required.

Including the mono- and dipole components in the Gibbs sampling chain,
it now reads
\begin{align}
\mathbf{w}_k^{i+1} &\leftarrow P(\mathbf{w}_k|\mathbf{d}_k, \mathbf{s}^{i}),\\
\mathbf{s}^{i+1}   &\leftarrow P(\mathbf{s}|C_{\ell}^{i}, \mathbf{d}, \mathbf{w}_k^{i+1}), \\
C_{\ell}^{i+1}     &\leftarrow P(C_{\ell}|\mathbf{s}^{i+1}).
\end{align}
The first step is computed as described in the previous paragraphs,
and the second step is computed by equation \ref{eq:meanfield}, with
the slight modification that the mono- and dipole contributions now
are subtracted from the data, $\mathbf{r}_k^{\textrm{s}} =
\mathbf{d}_k - \mathbf{T} \mathbf{w}_k$.

\subsection{Incomplete sky coverage}
\label{sec:incomplete_sky}

Perhaps the single most important complication in any CMB analysis is
proper treatment of foregrounds. With amplitudes up to several
thousand times the CMB amplitude, Galactic foregrounds will
necessarily compromise any cosmological result unless corrected and
accounted for. Unfortunately, there is currently a critical shortage
of robust component separation (or even just foreground removal)
methods, and the only reliable approach at the time of writing is
simply to mask out the most contaminated regions of the sky. On the
bright side, the flexibility in specifying foreground models that can
be implemented in the Gibbs sampling approach offers an attractive
avenue for progress. This will be explored further in future
publications. 

The Gibbs sampling approach supports two fundamentally different
methods for removing parts of the sky by means of a mask. First, the
most straightforward option from a conceptual point of view is simply
to set the inverse noise matrix to zero at all pixels within the
mask. This corresponds to saying that the noise level of these pixels
is infinite, and therefore that the data are completely
non-informative. No other modifications of the equations are
necessary. This is the solution chosen for the Commander
implementation.

However, this approach carries a considerable cost in the form of a
poorly conditioned coefficient matrix $\mathcal{A} = \mathbf{C}^{-1} +
\mathbf{A}^{\textrm{T}} \mathbf{N}^{-1}
\mathbf{A}$, which, as we will discuss
at greater length in the next section, results in slow convergence for
the conjugate gradient algorithm, and increased overall expense for
the Gibbs sampling. Recognizing this fact, an alternative approach was
chosen for the MAGIC implementation, namely to introduce a new
foreground component into the Gibbs sampling chain.

Let us recall the general sampling equation for the foreground
component \citep{wandelt:2004},
\begin{equation}
\begin{split}
\left(\mathbf{F}^{-1}_k+
  \mathbf{A}_k^{\textrm{T}}\mathbf{N}^{-1}_k\mathbf{A}_k\right)\,\mathbf{f}_k
  = \quad\quad\quad\quad\quad\quad\quad\quad\quad\quad \\
  =
  \mathbf{A}_k^{\textrm{T}} \mathbf{N}^{-1}_k
\mathbf{r}_k^{\textrm{fg}}+ \mathbf{F}^{-1/2}_k \omega_k^{(1)}+
\mathbf{A}_k^{\textrm{T}}\mathbf{N}^{-{\frac{1}{2}}}_k
  \omega_{k}^{(2)}. 
\end{split}
\label{foreground}
\end{equation}
Here $\mathbf{F}_k$ is the covariance matrix for the foreground prior,
$\mathbf{r}_k^{\textrm{fg}}$ is the residual map after removal of the
signal estimate and any other foregrounds already sampled by the
algorithm, and $\omega_k^{(i)}$ are vectors of uniform Gaussian
variates. Finally, $\mathbf{f}_k$ is the unconvolved foreground
sample.

For each pixel in the masked region, mainly the Galaxy but also some
point sources, we do not know the foreground contribution. The
maximally uninformative foreground prior for these pixels has infinite
variance. It corresponds to a complete lack of \emph{a priori}
knowledge of the foregrounds in the mask. By specifying maximal
ignorance of the foreground we allow the algorithm to determine the
level of the foreground in these pixels which is supported by the
data. Substituting this foreground prior into equation
\ref{foreground} creates a method to numerically marginalize over the
unknown foreground contribution in the masked pixels.

In the limit of 'infinite' variance, this sampling equation simplifies
to
\begin{equation}
\mathbf{A}_k\mathbf{f}_k =\mathbf{N}^{1/2}_k \omega_k
+\mathbf{r}_k^{\textrm{fg}}
\label{ivf}
\end{equation}
in the masked region and $\mathbf{f}_k=0$ outside. This is easy to
compute and avoids the use of the Conjugate Gradient solver, hence
saving computational time. 

With the introduction of this foreground component, the full Gibbs
chain reads
\begin{align}
\mathbf{f}_k^{i+1} &\leftarrow P(\mathbf{f}_k|\mathbf{d}_k, \mathbf{s}^{i}, \mathbf{w}_k^{i}),\\
\mathbf{w}_k^{i+1} &\leftarrow P(\mathbf{w}_k|\mathbf{d}_k, \mathbf{s}^{i}, \mathbf{f}_k^{i+1}),\\
\mathbf{s}^{i+1}   &\leftarrow P(\mathbf{s}|C_{\ell}^{i}, \mathbf{d},
\mathbf{f}_k^{i+1}, \mathbf{w}_k^{i+1}), \\
C_{\ell}^{i+1}     &\leftarrow P(C_{\ell}|\mathbf{s}^{i+1}).
\end{align}
Again, the only modifications in the two middle steps is a subtraction
of the foreground components from the corresponding residual maps,
$\mathbf{r}_k^{\textrm{s}} = \mathbf{d}_k - \mathbf{T} \mathbf{w}_k -
\mathbf{A}_k\mathbf{f}_k$ and $\mathbf{r}_k^{\textrm{md}} = \mathbf{d}_k -
\mathbf{A}_k (\mathbf{s} + \mathbf{f}_k)$.

As mentioned earlier, the main advantage of this approach is that the
uniform properties of the coefficient matrix $\mathcal{A}$ are
conserved, leading to a faster convergence for the conjugate gradient
solver, often reducing the number of iterations by a factor of
three. On the other hand, there is also a slight disadvantage in that
the correlations between consecutive Gibbs samples are stronger, since
information is carried over from sample to sample through the
foreground component. However, this is more than compensated by the
rapid CG convergence. We will return to these issues later.

\section{Computational considerations}

\subsection{Conjugate gradients and preconditioning}

As described in section \ref{sec:basic_gibbs}, equations
\ref{eq:meanfield} and \ref{eq:fluctmap} are the very heart of the
Gibbs sampling method, and its feasibility is directly connected to
our ability to solve those equations. For a low-resolution experiment
such as \emph{COBE}-DMR, which comprises a few thousand pixels or
multipole components, the system may be solved directly, for instance
through Cholesky decomposition. However, for a high-resolution
experiment such as \emph{WMAP}, with eight cosmologically important
maps of each several millions pixels, more sophisticated algorithms
must be employed, and the most efficient method currently available
for positive-definite matrices is the Conjugate Gradient (CG) method
\citep{golub:1996}. For a truly excellent review of this algorithm,
see \citet{shewchuk:1994}.

\begin{figure*}

\mbox{\subfigure[No sky cut, $\ell$-based ordering]{\epsfig{figure=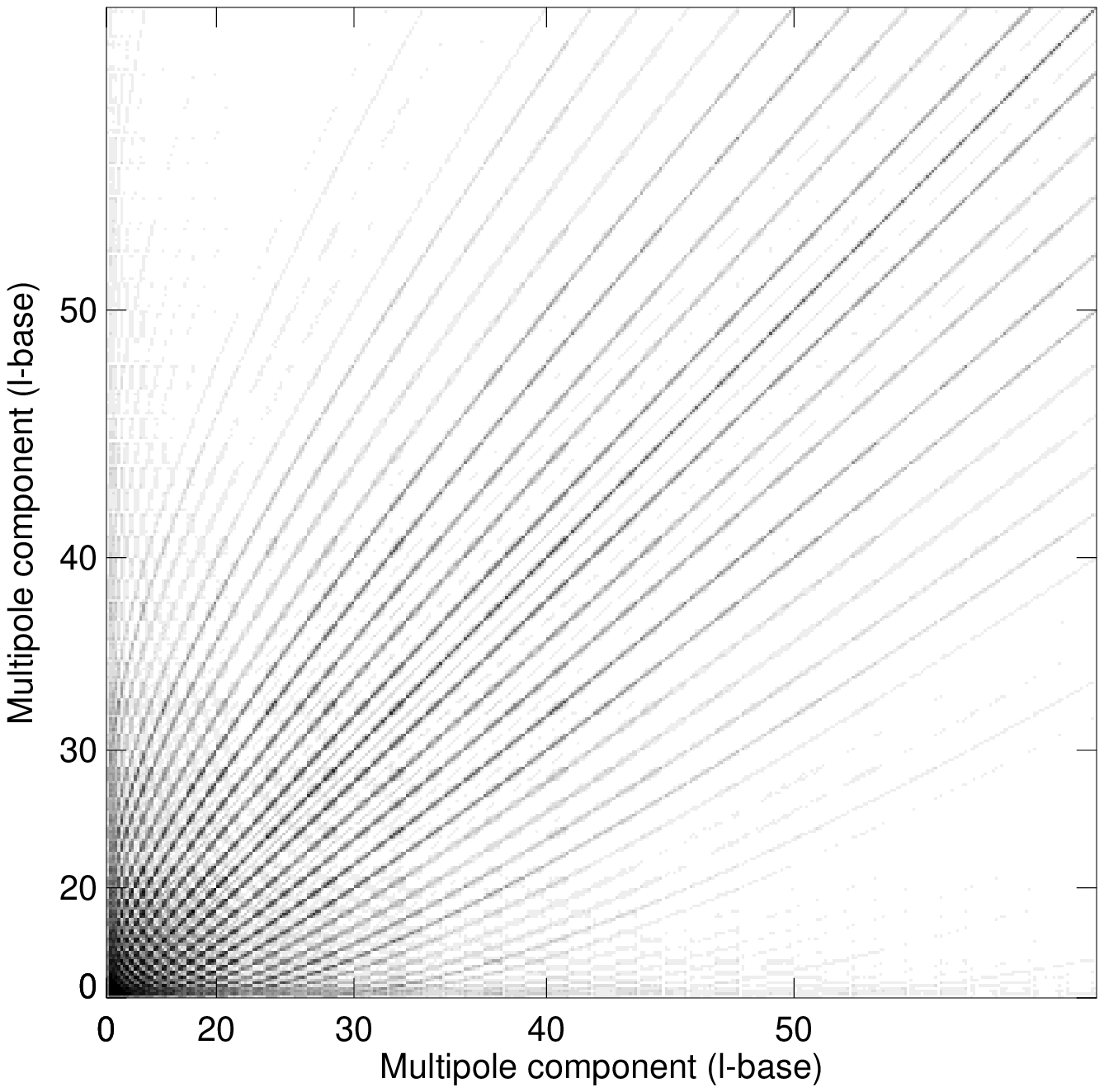,width=0.46\linewidth,clip=}}
      \quad
      \subfigure[No sky cut, $m$-based ordering]{\epsfig{figure=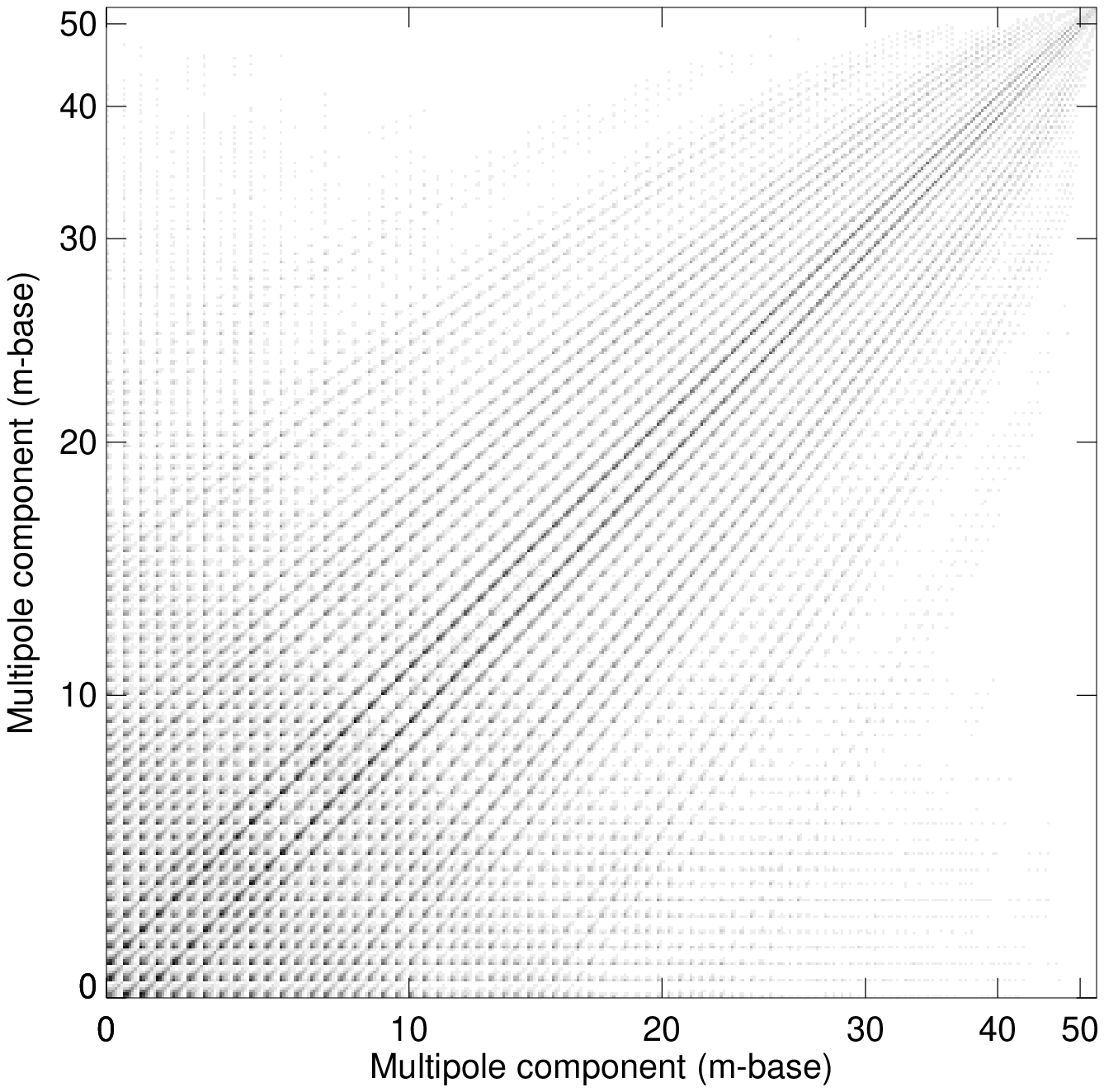,width=0.46\linewidth,clip=}}}

\mbox{\subfigure[Kp2 sky cut, $\ell$-based ordering]{\epsfig{figure=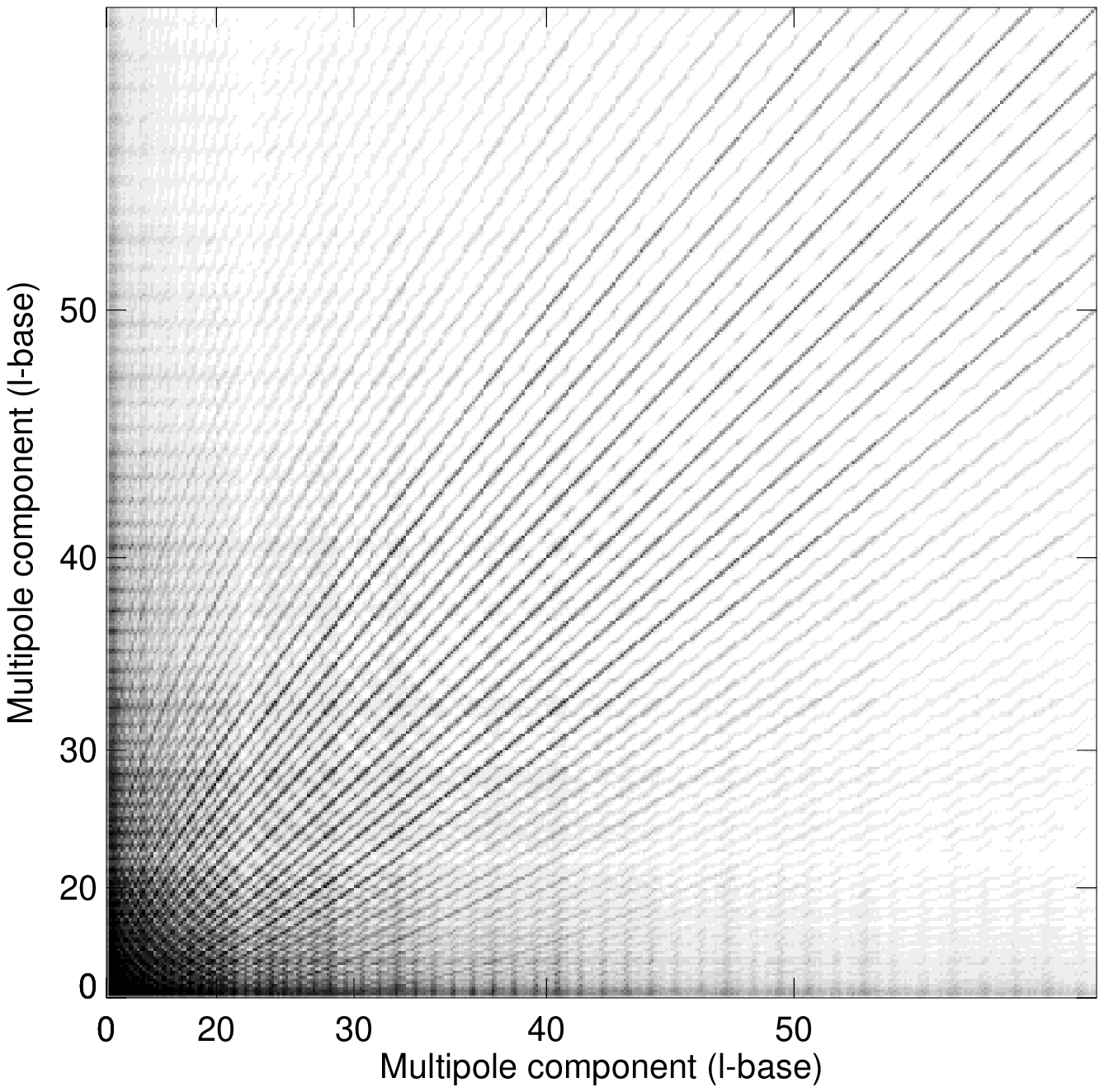,width=0.46\linewidth,clip=}}
      \quad
      \subfigure[Kp2 sky cut, $m$-based ordering]{\epsfig{figure=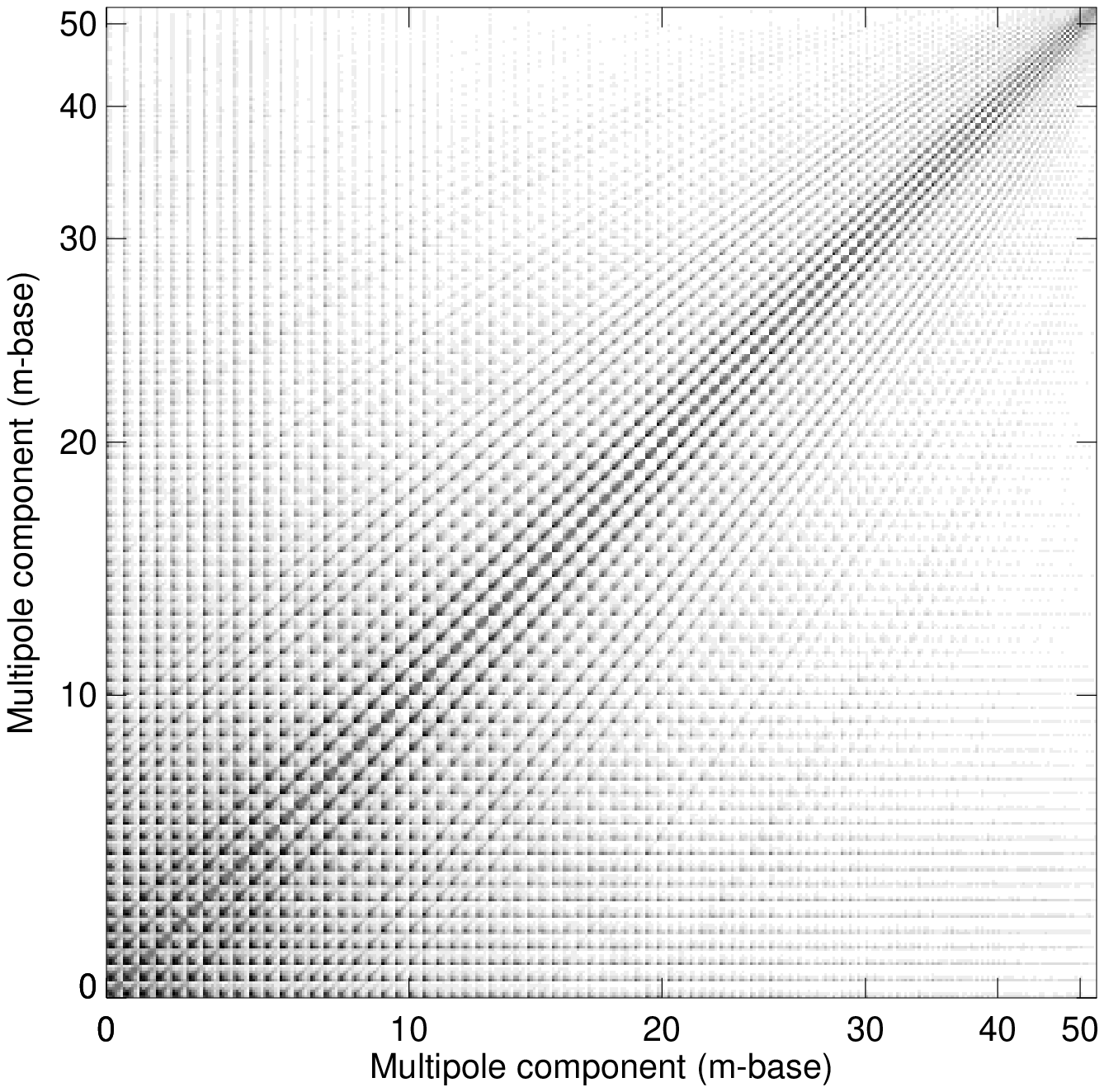,width=0.46\linewidth,clip=}}}

\caption{The coefficient matrices $\mathcal{A} = \mathbf{1} +
  \mathbf{C}^{1/2} [\sum_{k=1}^{N} \mathbf{A}_k^{\textrm{T}}
  \mathbf{N}^{-1}_k \mathbf{A}_k] \mathbf{C}^{1/2}$, summed over all
  eight \emph{WMAP} channels, and using the power spectrum estimated
  by the \emph{WMAP} team. All elements up to $\ell_{\textrm{max}} =
  59$ are included, a choice determined by plotting constraints
  only. The upper panels plot the matrix when the full sky is
  available, and the lower panels plot it when the Kp2 mask is
  applied. The elements are ordered by $\ell$-major (with pixel index
  $i$ given by $i = \ell^2 + \ell + m + 1$) in the left column, and by
  $m$-major in the right column. ($m$ increases as
  $m=0,-1,1,-2,2,\ldots$ from left to right, in steps of
  $\ell_{\textrm{max}}-|m|$. Within each $m$-block
  $\ell=|m|,|m|+1,\ldots,\ell_{\textrm{max}}$.) A solid black color
  indicates a signal-to-noise ratio larger than 5. }
\label{fig:sn_matrices}
\end{figure*}


The general problem is to solve a system of linear equations,
\begin{equation}
\mathcal{A}\mathbf{x} = \mathbf{b},
\label{eq:linsys}
\end{equation}
where the coefficient matrix $\mathcal{A}$ is very large.  In the case
of the first-year \emph{WMAP} data, $\mathcal{A}$ corresponds to a
system of three million equations in pixel space, and a system of
several hundred thousands equations in harmonic space (depending on
the $\ell_{\textrm{max}}$ of choice; see section
\ref{sec:correlations} for a discussion on how to choose an
appropriate $\ell_{\textrm{max}}$). Further, this coefficient matrix
is in general not sparse in either real space due to complicated
signal correlations, or in harmonic space due to complicated noise
correlations. However, favorable sparsity patterns may be obtained for
special scanning strategies and sky cuts \citep{oh:1999,wandelt:2003}.

For this reason, the sheer size of the problem poses a real problem,
and for many applications one may find that the system described above
is ill-conditioned. For instance, the solution vector $\mathbf{x}$
contains elements of very different magnitudes, and therefore
round-off errors can easily compromise the results. It is therefore
numerically advantageous to rewrite equations \ref{eq:meanfield} and
\ref{eq:fluctmap} as follows,
\begin{align}
\begin{split}
\left(\mathbf{1} + \mathbf{C}^{1/2}\left[\sum_{k=1}^{N}
  \mathbf{A}_k^{\textrm{T}} \mathbf{N}^{-1}_k
  \mathbf{A}_k\right]\mathbf{C}^{1/2}\right)\,\left(\mathbf{C}^{-1/2}\mathbf{x}\right)
   = \\ = \mathbf{C}^{1/2} \sum_{k=1}^{N} \mathbf{A}_k^{\textrm{T}}
  \mathbf{N}^{-1}_k \mathbf{r}_k^{\textrm{s}}
\end{split} \\
\begin{split}
\left(\mathbf{1} + \mathbf{C}^{1/2}\left[\sum_{k=1}^{N}
  \mathbf{A}_k^{\textrm{T}} \mathbf{N}^{-1}_k
  \mathbf{A}_k\right]\mathbf{C}^{1/2}\right)\,\left(\mathbf{C}^{-1/2}\mathbf{y}\right)
  = \\ = \mathbf{\omega}_0 + \mathbf{C}^{1/2}\sum_{k=1}^{N}
  \mathbf{A}_k^{\textrm{T}} \mathbf{N}^{-1/2}_k \mathbf{\omega}_k.
\end{split}
\label{eq:rewritten_eq}
\end{align}
The coefficient matrix $\mathcal{A} = \mathbf{1} + \mathbf{C}^{1/2}
\mathbf{A}^{\textrm{T}} \mathbf{N}^{-1} \mathbf{A}^{\textrm{T}}
\mathbf{C}^{1/2}$ is now much better behaved, and all elements of the
solution vector $\mathbf{C}^{-1/2}\mathbf{x}$ have unity
variance. Note also that the diagonal elements of $\mathcal{A}$ are
now simply the signal-to-noise ratios of the corresponding mode.

We choose to work in harmonic space in the following for several
reasons. First, in this space it is easy to limit the size of the
problem according to the signal-to-noise ratio of the data by choosing
an appropriate $\ell_{\textrm{max}}$. In pixel space one is always
forced to work with vectors of length $N_{\textrm{pix}}$. Second,
given the form of equations \ref{eq:meanfield} and \ref{eq:fluctmap},
two spherical harmonics transforms are eliminated by operating in
harmonic space in the first place, thereby reducing the total CPU time
by a factor of two. Finally, since we are mainly interested in the
power spectrum, an harmonic space based convergence criterion for the
CG search seems more natural than a pixel space based criterion.



One of the main advantages of the CG algorithm is that it does not
require inversion of the coefficient matrix, and we do not even need
to store it. All we need is the ability to multiply $\mathcal{A}$ with
a given vector $\mathbf{v}$, and solving a preconditioning
equation. We first consider the matrix multiplication operation. In
our setting, for which $\mathcal{A} = \mathbf{1} + \mathbf{C}^{1/2}
\mathbf{A}^{\textrm{T}} \mathbf{N}^{-1} \mathbf{A}^{\textrm{T}}
\mathbf{C}^{1/2}$, this is done in a step-wise fashion. First we
multiply each component $a_{lm}$ of the input vector by
$\sqrt{C_{\ell}} b_{\ell}$ (where $b_{\ell}$ is the product of the
beam and pixel window functions), and then we perform an inverse
spherical harmonic transform into real space. Here we multiply with
the inverse noise matrix, $N_{\textrm{obs}}/\sigma^2_0$, under the
assumption of uncorrelated noise. Then we perform an ordinary
spherical harmonic transform of the vector into harmonic space, where
we again multiply with the beam and square root of the power
spectrum. Finally we add the original vector. Thus, multiplication of
$\mathcal{A}$ is computationally equivalent to two spherical harmonic
transforms, and memory requirements are virtually
negligible\footnote{If accessible memory is sufficient on the
available computer, one may want to precompute the associated Legendre
polynomials, reducing the total CPU time typically by a factor of two
or three for \emph{WMAP} type maps in the current HEALPix
implementation. The memory requirement for doing so is $8
\;N_{\textrm{side}} \; \ell_{\textrm{max}}^2$ bytes, or on the order
of 1GB for $N_{\textrm{side}}, \ell_{\textrm{max}} \sim 512$,
$N_{\textrm{side}}$ being the HEALPix resolution parameter, which
corresponds directly to the number of pixels in the map through the
relation $N_{\textrm{pix}} = 12\, N_{\textrm{side}}^2$.}.

The efficiency of the CG algorithm is highly dependent on our ability
to construct a good preconditioner (e.g., Oh et al.\ 1999), and two
preconditioners have been proposed for this problem so far, both
approximating $\mathcal{A}^{-1}$ in harmonic space.  First, under the
assumption of white, but non-uniform, noise, the inverse real-space
noise covariance matrix may be written as a simple inverse noise rms
map, $N^{-1}(\theta, \phi)$, which again may be expanded into
spherical harmonics,
\begin{equation}
N^{-1}(\theta,\phi) = \sum_{\ell,m} a_{\ell m} Y_{\ell m}(\theta,\phi).
\end{equation}
The inverse noise matrix in spherical harmonic space is then \citep{hivon:2002}
\small
\begin{equation}
\begin{split}
\mathbf{N}^{-1}_{\ell_1 m_1, \ell_2 m_2} = &\sum_{\ell_3, m_3} a_{\ell_3 m_3} (-1)^{m_2}
\biggl[\frac{(2\ell_1 + 1)(2\ell_2 + 1)(2\ell_3 + 1)}{4\pi}\biggr]^{1/2} \\
&\times \left(
\begin{array}{ccc}
\ell_1 & \ell_2 & \ell_3 \\
0 & 0 & 0
\end{array}
\right)
\left(
\begin{array}{ccc}
\ell_1 & \ell_2 & \ell_3 \\
m_1 & -m_2 & m_3
\end{array}
\right)
\end{split}
\end{equation}
\normalsize

A very simple preconditioner may therefore be defined in terms of the
diagonal elements only,
\begin{equation}
\mathbf{M}_{\ell_1 m_1, \ell_2 m_2} = \biggl[1 + C_{\ell_1}
b^2_{\ell_1} \mathbf{N}^{-1}_{\ell_1 m_1, \ell_2 m_2} \delta_{\ell_1
\ell_2}\delta_{m_1 m_2}\biggr]^{-1}.
\end{equation}
While satisfactory for the simplest applications, we find that it
takes about 300 iterations to solve for the \emph{WMAP} data
consisting of all eight cosmologically interesting bands with this
preconditioner (applying the Kp2 mask directly), making the total
solution of the problem very expensive.

By considering the overall structure of the inverse noise matrix,
\citet{oh:1999} proposed to use a block-diagonal matrix. In the limit
of perfect azimuthal symmetry of both the galactic cut and the noise
distribution, $\mathbf{N}^{-1}$ is orthogonal with respect to $m$, and
therefore it makes sense to also include all elements having $\ell_1
\ne \ell_2, m_1 = m_2$ up to some arbitrary limit $m_{\textrm{max}}$.
At higher $m$'s, the diagonal preconditioner is used. \citet{oh:1999}
claims to achieve convergence in six iterations with this
preconditioner for properties corresponding to the two-year
\emph{WMAP} data, but, unfortunately, we have not yet been able to
reproduce this performance. From our experiments it seems the
combination of a highly non-symmetric Kp2 cut, 700 resolved point
source cuts, and a noise distribution tilted with respect to the
galactic plane introduces significant couplings between different
$m$'s.

In figure \ref{fig:sn_matrices} we have plotted the coefficient
matrices corresponding to the first-year \emph{WMAP} data in two
different orderings, both $\ell$-major and $m$-major (see caption for
details), and with and without application of the Kp2 mask. In the
limit of uniform noise and no galactic cut, these matrices would all
be diagonal, and convergence would be reached in one single CG
iteration using even the diagonal preconditioner.

However, as seen in the top two panels of figure
\ref{fig:sn_matrices}, adding non-uniform noise to the problem
introduces significant coupling between different modes, which again
leads to poorer CG performance. In the left panel, we see that the
largest absolute values are found at low $\ell$'s, which of course is
not very surprising, considering that these matrices are a measure of
the signal-to-noise ratio. In the right panel we see the same matrix
organized as $m$-major, and in the limit of azimuthal symmetry, this
would be a strictly block-diagonal matrix with very small block
elements. The preconditioner proposed by \citet{oh:1999} consists of
the inverses of those blocks. But, as we see, there are many
off-diagonal elements in this matrix, and, indeed, the dominant
elements actually seem to be components for which $|m_1 - m_2| =
1$. However, if we had computed these quantities in the ecliptic
frame, rather than in the galactic, then the matrix is likely to be
dominated by the $m_1 = m_2$ elements, and possibly even by the $m =
0$ elements.

The bottom two panels show a similar set of matrices, but in this case
the Kp2 mask has been applied to the sky. And, as mentioned in section
\ref{sec:incomplete_sky}, this has the highly undesirable effect of
magnifying the off-diagonal elements through mode-to-mode coupling
considerably. Unfortunately, neither of these matrices have a very
dominant symmetry structure, and it is therefore difficult to
establish an optimal preconditioner.

Nevertheless, based on the structures seen in the lower left panel in
figure \ref{fig:sn_matrices} a third alternative was chosen for the
Commander implementation. Rather than including only the diagonal
elements, or only $m_1 = m_2$ elements as \citet{oh:1999} do, we
include \emph{all} elements up to some arbitrary $\ell_{\textrm{max}}$
(typically $\ell_{\textrm{max}}\approx 50$--70 for \emph{WMAP}), and
at higher $\ell$'s we include only the diagonal elements. The
required memory requirements for this matrix scales as
$\mathcal{O}(\ell_{\textrm{max}}^4)$, and are thus quite expensive,
but in practice, the real limitation is the CPU time required for its
Cholesky decomposition (which scales as
$\mathcal{O}[\ell_{\textrm{max}}^6]$) rather than memory requirements
for its storage. For $\ell_{\textrm{max}} = 50$, the memory
requirements are 52 MB and the CPU time for Cholesky decomposition is
on the order of one or two minutes. Obviously, the latter number must
be compared to the CPU time it takes to perform one CG iteration and
the number of iterations saved. And yet, even with this rather
expensive preconditioner, we find that the CG search converges in
about 60 iterations for the combined first-year \emph{WMAP} data and a
norm-based fractional convergence criterion of $10^{-6}$. Thus, our
performance is not as impressive as the six iterations achieved by
\citet{oh:1999}. Work on this issue is still on-going, and a hybrid
of all three variants may prove to be the ultimate solution.

In contrast to the Commander implementation, MAGIC does not apply a
sky cut directly, but instead it introduces a new random field into
the sampling chain. The appropriate coefficient matrix is therefore
the one shown in the upper left panel of figure
\ref{fig:sn_matrices}. This choice has a very positive effect in terms
of CG performance, and one routinely achieves convergence within 20
iterations using just the simple diagonal preconditioner for a
first-year \emph{WMAP} type experiment. However, as we will see later,
the cost for this performance comes in the form of a slightly longer
correlation length in the Markov chain, and therefore fewer
independent samples.

\subsection{Parallelization}
\label{sec:parallelization}

The main limitation for the Gibbs sampling method is CPU time. Even
though the scaling of the method is equivalent to that of a spherical
harmonics transform for a \emph{WMAP} type analysis, one has to
perform this operation many times, and the total prefactor of the
algorithm is therefore large. Specifically, the number of spherical
harmonic transforms to produce one Gibbs sample is two times the
number of CG iterations, times the number of frequency bands. The
total number of transforms for computing one sample from an eight-band
\emph{WMAP} data set is then typically on the order of 1000 for the
Commander approach (reaching convergence in 60 iterations) and 350 for
the MAGIC approach (reaching convergence in 20 iterations). Knowing
that one harmonic transform takes about 5 seconds for
$N_{\textrm{side}} = 512$ and $\ell_{\textrm{max}} = 512$, the total
CPU time required for one single Gibbs sample is therefore on the
order of one or two hours for Commander and half an hour for
MAGIC. Obviously, parallelization is essential to produce a sufficient
number of samples.

\begin{figure}

\mbox{\epsfig{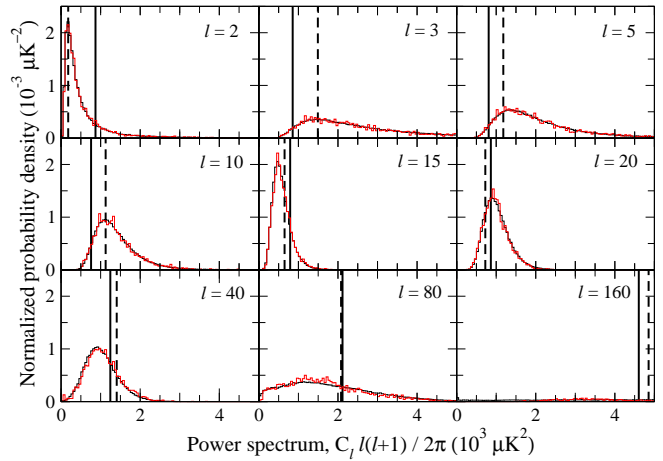}}

\caption{A few selected histograms of the power spectrum samples
  produced in the low-resolution analysis. The black histograms are
  generated by Commander, and the red histograms are generated by
  MAGIC; the agreement is striking, demonstrating that both codes work
  as expected. The vertical solid lines indicate the theoretical input
  spectrum, and the dashed lines indicate the realization specific
  spectrum. The agreement between the peak position of the histograms
  and the dashed lines is excellent at low $\ell$'s. At high $\ell$'s,
  however, the distributions are completely dispersed, reflecting the
  noise domination in this regime.}
\label{fig:sim_hist}
\end{figure}

Two fundamentally different approaches may be taken in this respect.
Either one may choose to run one single Markov chain and parallelize
the spherical harmonic transforms internally. Since the HEALPix
routines operate on pixel rings of constant latitude, this can be done
quite efficiently by letting each processor compute its own ring.
Nevertheless, optimal speed-up will not be achieved, and the
implementation will be somewhat complicated.

\begin{figure*}

\mbox{\epsfig{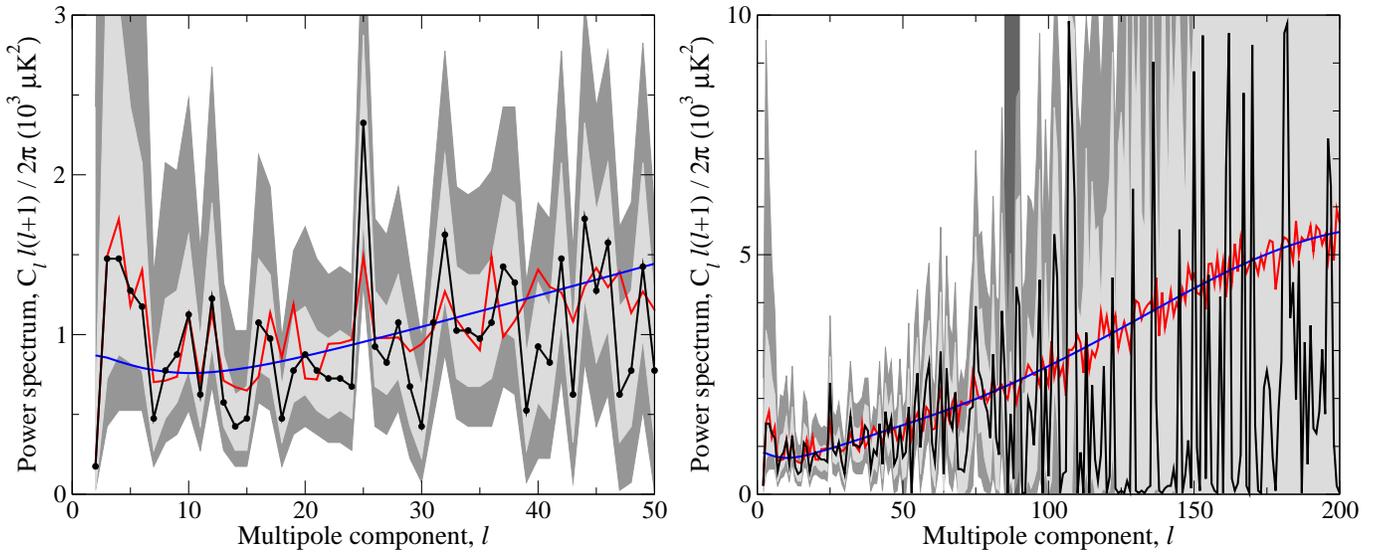}}

\caption{Power spectrum results from the low-resolution simulations.
  The blue line indicates the theoretical spectrum from which a random
  Gaussian realization was drawn, and the red curve is the power
  spectrum of that particular realization. The black curve shows the
  marginalized estimates of this spectrum produced by Commander, while
  the gray bands indicate the 1 and $2\sigma$ confidence bands. The
  information in the two panels is the same, but for different ranges
  in $\ell$. The dark gray vertical region in the right panel
  indicates where the signal-to-noise ratio is approximately unity.}
\label{fig:sim_spectrum}
\end{figure*}

The other approach is to take advantage of the fact that this method
is truly a Monte Carlo method, and one can therefore let each
processor run its own Markov chain. The most important advantages of
this approach are optimal speed-up and the possibility to initialize
each chain with a different first guess. As we will see in the next
section, consecutive Gibbs samples in the Markov chain are highly
correlated in the low signal-to-noise regime, and producing a larger
number of independent samples is therefore quite expensive. If we
have a rough approximation of the true spectrum and its uncertainties
(as we usually do, through a MASTER type analysis; Hivon et al.\
2002), we can partially remedy this problem by initializing each
Markov chain with an independent power spectrum.

The major drawback of this latter parallelization scheme, however, is
that each Markov chain will necessarily be quite short, perhaps only
twenty to fifty samples. This problem is due to the fact that most
current super-computer facilities have a maximum wall-clock time limit
of 24 to 72 hours, and therefore the maximum length of one chain is on
the same order of magnitude. Of course, one may store intermediate
results and restart the computations after every cycle, but this only
increases the total length by a factor of a few, not by hundreds.

We have chosen a combination of external and internal parallelization
in our implementation, by recognizing the fact that we will in general
be analyzing multi-frequency data sets consisting of
$N_{\textrm{band}}$ maps. We may therefore let $N_{\textrm{band}}$
processors work on the same Markov chain, each processor transforming
one band. Thus, optimal speed-up is not compromised, while the length
of the chains is increased by the same factor. In future versions we
will also implement fully internal parallelization in the HEALPix
routines map2alm and alm2map, to have the option of focusing all the
computational resources into one single chain.

\section{Simulations}

In this section we apply the Commander and MAGIC codes to simulated
data sets for which all components are perfectly known. The goals are
two-fold. Firstly, we wish to demonstrate that the codes produce
results consistent with theoretical expectations, and secondly, we
seek to gain insight on what limitations of the algorithm we can
expect to meet in real-world applications, when CPU time is limited.

A number of different simulations are analyzed in the following
sections, each designed to highlight some specific feature. First, in
order to establish the asymptotic behavior of the algorithm, we study
a data set of smaller size than the full-resolution \emph{WMAP}
data. Specifically, we construct a data set at intermediate resolution
($N_{\textrm{side}} = 128$; 196\,608 pixels), for which the CPU time
per sample is on the order of 10 seconds. Thus, CPU time is not a
dominating problem, and we can establish the Markov chain correlation
lengths and power spectrum correlation matrix to great accuracy. The
burn-in time is also considered.

Finally, we make two simulations at full \emph{WMAP} resolution in
order to confirm that the overall results from the low-resolution
analysis carry naturally over to higher resolutions. This time the CPU
cost is the limiting factor, and the main goal of this section is in
fact to demonstrate that the Gibbs sampling method is able to handle
even large data sets, such as the \emph{WMAP} data. This analysis
mimics the analysis of the first-year \emph{WMAP} data presented by
\citet{odwyer:2004}, in that it is run on a super-computer with many
short, parallel chains. The only difference between the two runs is
that either white or correlated noise are added to the CMB
simulations. This way we test whether the assumption of white noise
may compromise the scientific results in the presence of small, but
non-negligible, noise correlations. We find that this is not a
significant problem for the first-year \emph{WMAP} data.

\subsection{Low-resolution simulations}

The main goal of the low-resolution simulations is to study the
asymptotic behavior of the method when the number of independent
samples is very high. On the one hand, this allows us to verify that
the codes work as expected without worrying about errors introduced
because of a limited number of samples, and on the other, essential
quantities such as the Markov chain correlation length and the power
spectrum correlation matrix may be established to a high degree of
accuracy.

In order to facilitate such long-chain analyses, we study maps with
relatively low resolution, $N_{\textrm{side}} = 128$, but with
properties corresponding to a consistently down-scaled
\emph{WMAP}-type experiment. Specifically, we generate a CMB sky from
the best-fit \emph{WMAP} running index spectrum, and convolve this sky
with modified version of the \emph{WMAP} beams. The beams are made four times
wider by replacing their original Legendre transform with $b_{\ell}
\rightarrow b^{\textrm{low res}}_{\ell} = 1/4 \; \sum_{\ell'=0}^{3}
b_{4\ell+\ell'}$.

The noise components are generated by degrading the original
\emph{WMAP} noise rms maps\footnote{The noise rms maps are defined by
$\sigma_i(p) = \sigma_{0}^i/\sqrt{N_{\textrm{obs}}^i(p)}$, where
$\sigma_{0}^i$ is the average sensitivity of the various bands, and
$N_{\textrm{obs}}^i(p)$ is the number of observations for each pixel
$p$.}  to $N_{\textrm{side}}=128$ by simple averaging over pixels in
the HEALPix nested organization. Thus, the noise per low-resolution
pixel in our simulated maps is about the same as that for each
high-resolution pixel in the full-sized \emph{WMAP} data. The
signal-to-noise ratio is therefore downscaled to the appropriate
resolution, a fact which will be important when studying the
relationship between the correlation length and the signal-to-noise
ratio.

\begin{figure*}



\mbox{\subfigure[]{\epsfig{figure=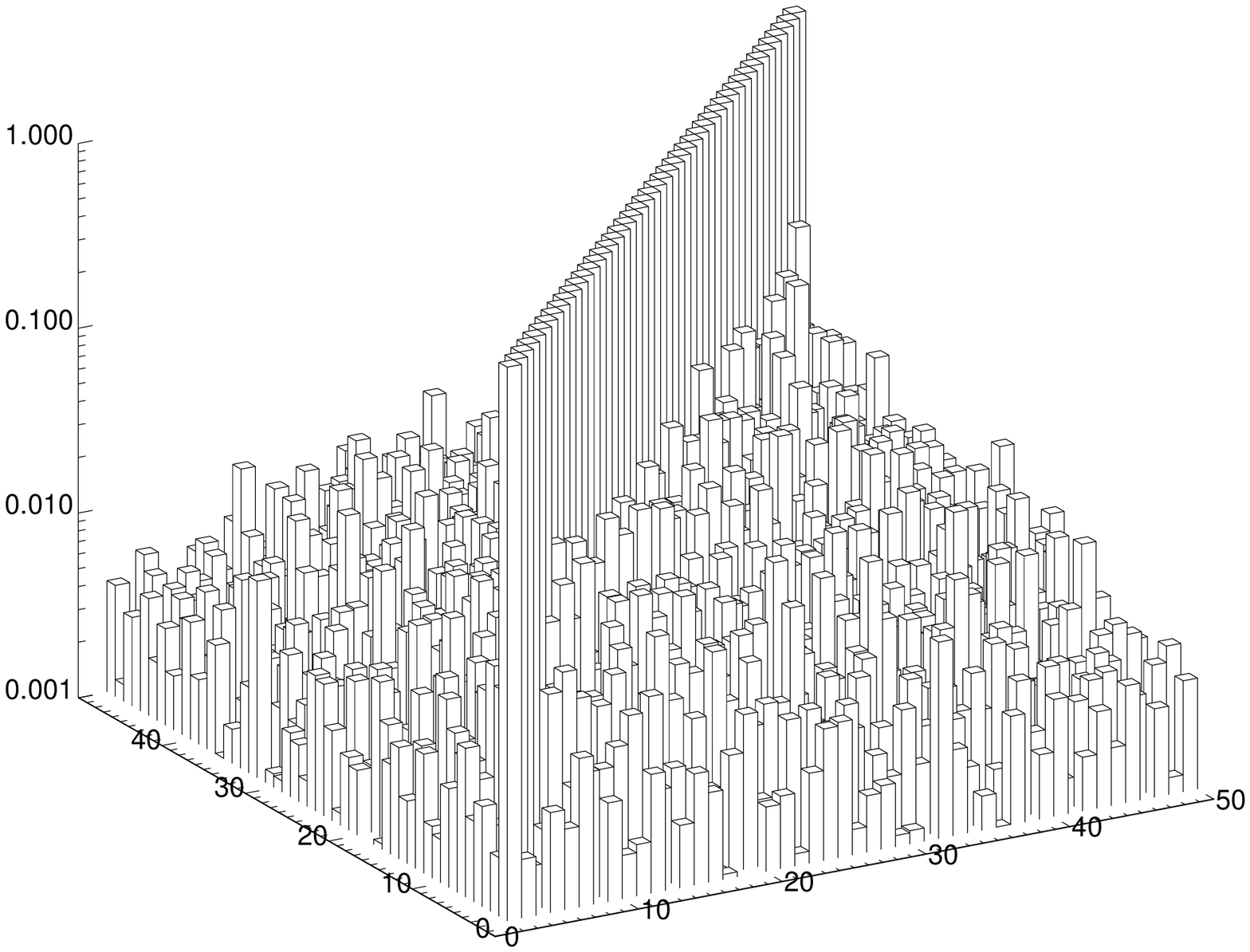,width=0.46\textwidth,clip=}} \\
      \subfigure[]{\epsfig{figure=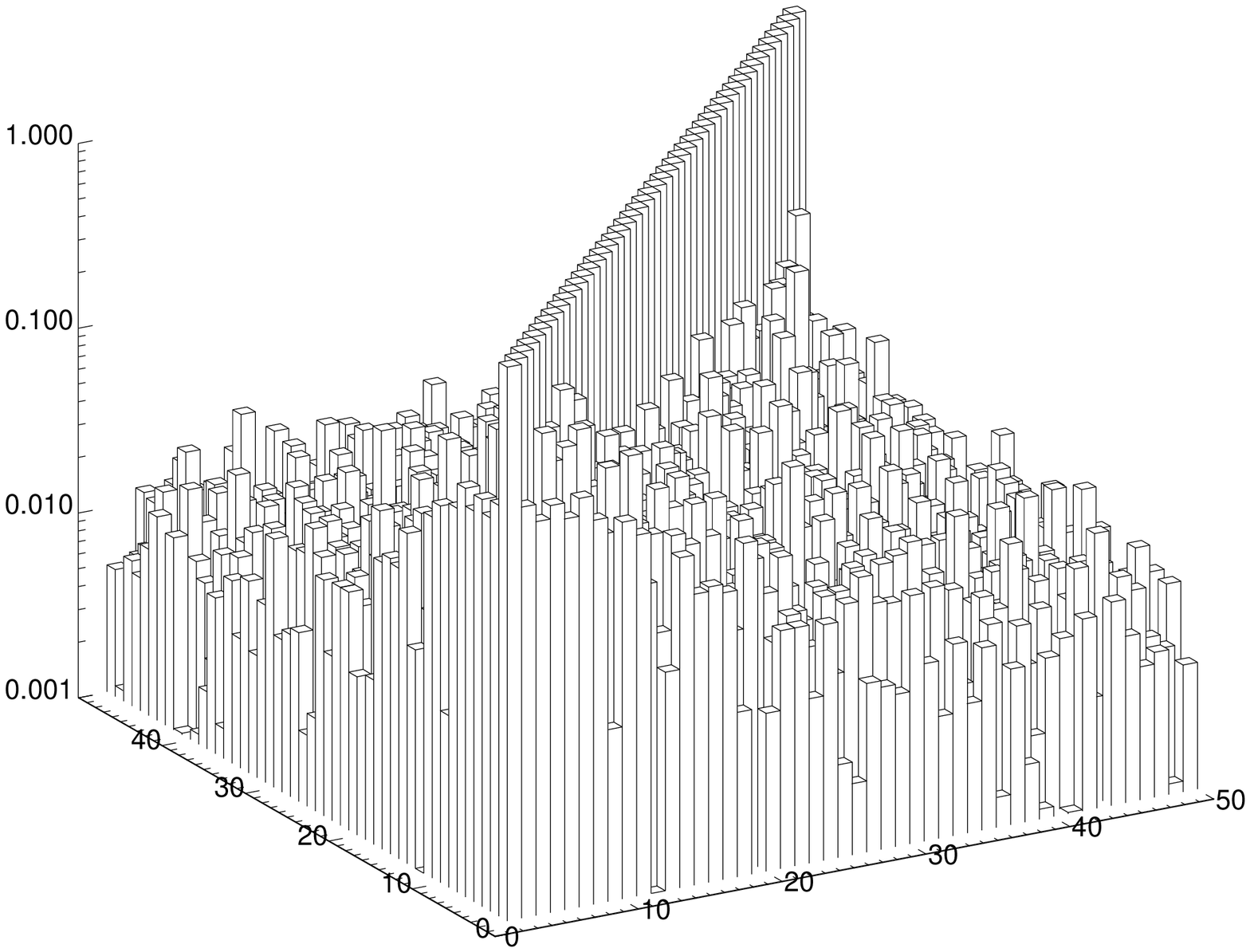,width=0.46\textwidth,clip=}}}

\caption{The absolute value of a) the $C_{\ell}$ correlation matrix,
  and b) the correlation matrix of the power spectra computed from the
  sampled maps, $\mathbf{s}$. The latter matrix may in many respects
  be interpreted as the harmonic space mode-mode coupling matrix.}
\label{fig:covarmat}
\end{figure*}

We also want to study the effect of residual mono- and dipoles on the
cosmological power spectrum, and we therefore add a random mono- and
dipole contribution with an artificially large amplitude (on the order
of tens to a hundred mK) to the signal plus noise map. The
reconstructed values are then later compared with the exact input
values.

\begin{figure}

\mbox{\epsfig{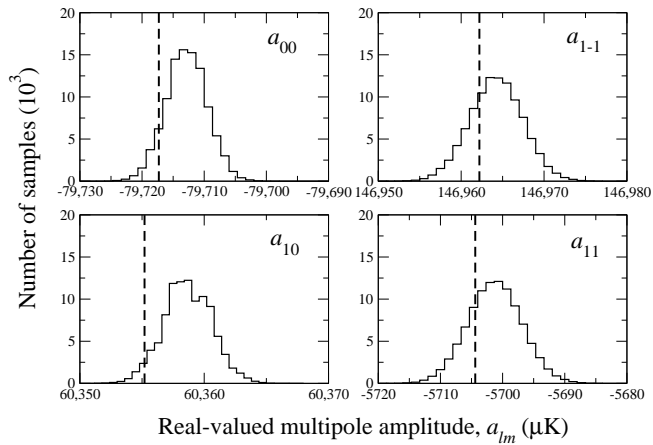}}

\caption{Distributions of mono- and dipole samples. The dashed,
  vertical lines show the true input value, and the histograms show
  the sampled values. The observed shift between the mode of the
  distributions and the input values is most likely due to internal
  couplings between the components. The most important result,
  however, is that the scatter is very small, with typical rms values
  smaller than $10\; \mu \textrm{K}$, even for the unrealistically
  large input values used in this experiment.}
\label{fig:md_results}
\end{figure}

Finally, we generate a degraded mask to match this resolution, based
on the \emph{WMAP} Kp2 mask as defined by \citet{bennett:2003b}. This
mask is downgraded to $N_{\textrm{side}}=128$ by requiring that all
high-resolution sub-pixel within a $N_{\textrm{side}}=128$ pixel
(again, in the HEALPix nested organization) are included by the
original mask. Thus, this mask is very slightly expanded compared to
the actual Kp2 mask.

\subsubsection{Verification of algorithms and codes}

In the first test, we apply the Commander and MAGIC codes to one
single band from the data set described above, namely to the V1
band. Commander was run for 100\,000 samples, while MAGIC was run for
4000, with the main goal of comparing the codes, verifying that they
produce identical output.

The results from this exercise are shown in figure
\ref{fig:sim_hist}. The black probability densities show the Commander
results, while the red histograms show the MAGIC results. The
agreement is striking, and this is a strong confirmation that the
codes work as expected, and that the minor differences in
implementational details discussed earlier do not affect the
scientific results.

The dashed lines show the true, underlying CMB power spectrum value,
which should theoretically coincide with the peaks of the histograms,
in the limit of full sky coverage and no noise. At low $\ell$'s, we
see that this is indeed the case. Here it is also worth recalling that
we added artificially large mono- and dipole components to the
simulations (several orders of magnitudes larger than what is
realistic), and this does still not compromise the results.

In figure \ref{fig:sim_spectrum} we have plotted the full spectrum
computed from the 100\,000 sample run. The input spectrum is marked in
red, the ensemble-averaged spectrum in blue, and the maximum
likelihood solution from the Gibbs sampler in black. The gray bands
indicate 1 and $2\sigma$ confidence regions for the power spectrum. 

At low $\ell$'s the discrepancy between the estimated and input
spectra is primarily due to the galactic cut, while at high $\ell$'s
it is primarily due to noise. In particular, we see that the maximum
likelihood estimate actually drops to zero for many of the low
signal-to-noise ratio modes, which, again, is the expected behavior
for a maximum likelihood estimator in the noise-dominated regime.

In figure \ref{fig:covarmat} we plot two different correlation
matrices, each on the form
\begin{equation}
C_{\ell \ell'} = \biggl<
\frac{C_{\ell}-\bigl<C_{\ell}\bigr>}{\sqrt{\textrm{Var} \,C_{\ell}}}
\frac{C_{\ell'}-\bigl<C_{\ell'}\bigr>}{\sqrt{\textrm{Var} \,C_{\ell'}}} \biggr>.
\end{equation}
The averages are taken over the 100\,000 samples in the Markov chain
described above. The left panel shows the correlation matrix of the
sampled power spectra, which are basically uncorrelated by
construction, while the right panel shows the correlation matrix of the
power spectra, $\sigma_{\ell}$, computed from the sampled maps,
$\mathbf{s}$. The latter matrix is related to the correlation matrix of
the maximum likelihood power spectrum found by maximizing the
posterior, and mainly describes mode-mode coupling due to the cut sky
on these scales. 

Finally, in figure \ref{fig:md_results} we plot the distributions of
the mono- and dipole samples and compare them to the input values,
marked by dashed, vertical lines. Although there certainly is a
discrepancy between the distribution modes and the input values, the
overall rms values are very small, on the order of 5--$10\;\mu
\textrm{K}$, and consistent with the fluctuation level expected for a
single realization. Further, there is some coupling between the mono-
and dipole modes due to the galactic cut, which could be
important. However, since our sole interest in these components lies
in removing them, rather than estimating them, this is not an
important problem for our purposes. In fact, given the very small
impact of these very large mono- and dipole components, we feel
confident that the cosmological low-$\ell$ spectrum is not compromised
by mono- and dipole issues.

\subsubsection{Convergence and correlations}
\label{sec:correlations}

We now turn to the issues of convergence, correlation length and
burn-in time, all of which must be thoroughly understood in order to
design and optimize a real-world analysis properly. The problem can
be plainly stated as follows: How many Gibbs samples do we need to
estimate the power spectrum with sufficient accuracy in order to be
limited by non-algorithmic issues?  As we will see, the answer depends
intimately on which angular scales we wish to consider, a conclusion
which is most easily seen by going back to the Gibbs sampling scheme.

The algorithm works as follows: First we assume some arbitrary (but
hopefully reasonable) power spectrum, and compute a Wiener-filtered
map based on that spectrum. Then we add a fluctuation term which
replaces the power lost both to noise and to the galactic cut. The
sum of those two terms mimics a full-sky, noiseless map with a power
spectrum determined by the data in the high signal-to-noise regime,
and by the assumed power spectrum in the low signal-to-noise regime.
From this full-sky power spectrum we then draw a new spectrum, which
subsequently is taken as the input spectrum for the next Gibbs
iteration.

\begin{figure*}
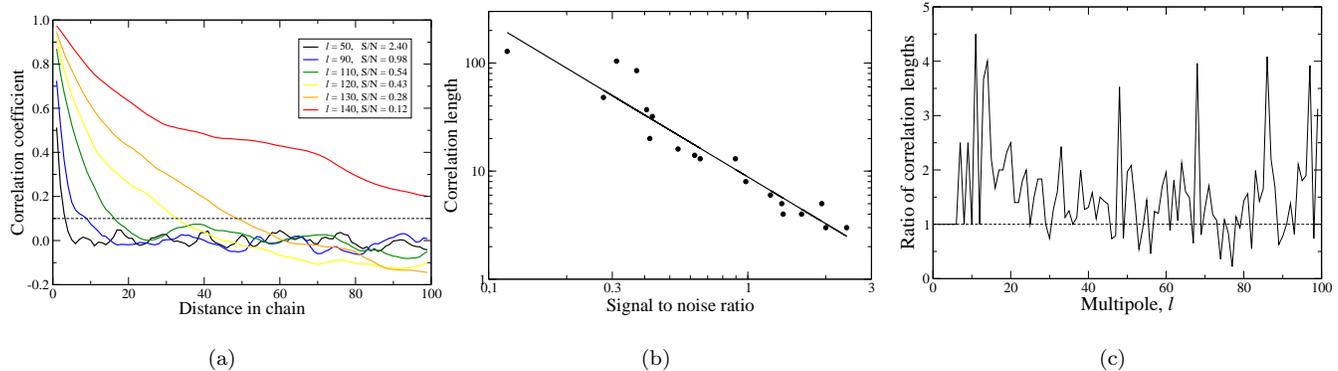




\mbox{\subfigure[]{\label{fig:sim_corlength_a}\epsfig{figure=f8a.eps,width=0.32\textwidth,clip=}}\\
      \subfigure[]{\label{fig:sim_corlength_b}\epsfig{figure=f8b.eps,width=0.32\textwidth,clip=}} \\
      \quad\subfigure[]{\label{fig:sim_corlength_c}\epsfig{figure=f8c.eps,width=0.32\textwidth,clip=}}}

\caption{The relationship between the signal-to-noise ratio and the
  correlation length of the Markov chains. a) The correlation
  function of the Markov chain, computed for a few selected multipoles
  from a Commander chain. Note how the correlations are stronger when
  the signal-to-noise ratio decreases. b) The typical correlation
  length as a function of signal-to-noise ratio. The typical
  correlation length is defined as the distance for which the
  correlation functions in the left panel drop below 0.1. c) The ratio
  of the MAGIC correlation length to the Commander correlation length,
  as a function of multipole.}
\label{fig:sim_corlength}
\end{figure*}

The crucial point is that the random step size in the final stage is
determined by the cosmic variance alone. Our goal is to probe the full
probability distribution which includes both noise \emph{and} cosmic
variance. In the high signal-to-noise regime, the difference does not
matter. Sequential Gibbs samples are therefore for all practical
purposes uncorrelated. The opposite is true in the low signal-to-noise
regime: since the distance between the two samples is determined by
the cosmic variance, while the full distribution is dominated by the
much larger noise variance, two sequential samples will be strongly
correlated.

This problem is a severe limitation for the Gibbs sampling technique
in its current formulation. It makes it very expensive to probe the
low signal-to-noise regime completely. The Gibbs sampling technique is
only a special case of the more general Metropolis-Hastings
framework. Other sampling schemes may be devised which break the
correlation between neighboring samples. This will be the topic of a
future publication, and for now our main goal is to quantify this
effect, rather than eliminate or minimize it.

We take advantage of the low-resolution simulations in order to
quantify these correlations. Specifically, we consider the power
spectrum values at constant $\ell$ in the Markov chain as independent
functions, and study the correlations in these chains as a function of
$\ell$. The statistic we choose for this study is a simple
auto-correlation function,
\begin{equation}
C(n) =
\biggl<\frac{C_{\ell}^i - \bigl< C_{\ell} \bigr>}{\sqrt{\textrm{Var} \,C_{\ell}}}\frac{C_{\ell}^{i+n}
  - \bigl< C_{\ell}\bigr>}{\sqrt{\textrm{Var} \,C_{\ell}}}\biggr>.
\end{equation}
Here $n$ is the distance in the chain measured in number of
iterations. Such functions are plotted in figure
\ref{fig:sim_corlength_a} for six different $\ell$'s, computed from a
new Commander chain consisting of 3800 samples, including all eight
bands.

As expected, the correlations become stronger as $\ell$ increases, or,
equivalently, as the signal-to-noise ratio decreases. In this
particular case, the signal-to-noise ratio is unity at approximately
$\ell=85$, and therefore the spectrum is limited by cosmic variance at
smaller $\ell$'s. This translates into a very short correlation
length for $\ell=50$ in this case, and consequently into a high
efficiency in terms of independent samples. On the other extreme, the
correlation length at $\ell=140$ is very, very long, and with only
3800 samples in the chain, we have only a very few independent samples
from which to form our power spectrum estimate.

We can take this exercise one step further and define a typical
correlation scale for each $\ell$, by computing the scale at which the
correlation function drops under, say, 0.1. In figure
\ref{fig:sim_corlength_b} we have plotted this correlation length
directly as a function of the signal-to-noise ratio, and from this
plot there seems to be a well-defined relationship between these two
quantities. In fact, we will use this relation to estimate how many
samples we need in the actual \emph{WMAP} analysis later on. For now
we note that with 3800 samples, as in the above case, we have about
200 independent samples at a signal-to-noise ratio of 0.6, which
corresponds to $l\approx 105$. In other words, it would be rather
optimistic to believe in the power spectrum based on these samples at
$\ell$'s higher than, say, 110.

Another lesson to be learned from these plots is that the correlation
length increases very rapidly with $\ell$, once entering the low
signal-to-noise regime. This is an important point to realize when
desiging a new analysis: probing the low signal-to-noise regime with
the current implementation of the Gibbs sampling algorithm is
extremely expensive. It may therefore often be desirable to limit
$\ell_{\textrm{max}}$ to the $\ell$ corresponding to a signal-to-noise
ratio of, say, 0.5 or 0.25. The saved CPU time\footnote{The algorithm
scales overall as $\mathcal{O}(\ell_{\textrm{max}}^3)$.} may then be
spent on producing more independent samples in the high and
intermediate signal-to-noise regimes. However, truncating the system
this way does modify the global solution, and care must therefore be
taken with respect to the highest $\ell$'s. In general, the larger the
sky cut, the more high-$\ell$ modes will have to be discared from the
final power spectrum, since mode-mode couplings spread the sharp
$\ell$-space cut-off into a wide range of multipoles. In practice, it
is convenient to pre-define some range of $\ell$'s of interest, and
then increase $\ell_{\textrm{max}}$ until that range becomes stable.

In figure \ref{fig:sim_corlength_c} we have plotted the ratio of the
MAGIC correlation length to the Commander correlation length, and here
it is seen, as noted earlier, that the MAGIC correlation length is
typically a factor of 1.5--2 longer at low $\ell$'s, resulting in a
smaller number of independent samples of the same factor. Of course,
this is both caused and made up by the fact that MAGIC handles the
incomplete sky coverage differently than Commander. Since MAGIC
obtains convergence in the CG search roughly three times faster than
Commander (using a very crude preconditioner), the codes do perform
quite similarly in terms of total CPU time per independent sample.

\begin{figure}

\mbox{\epsfig{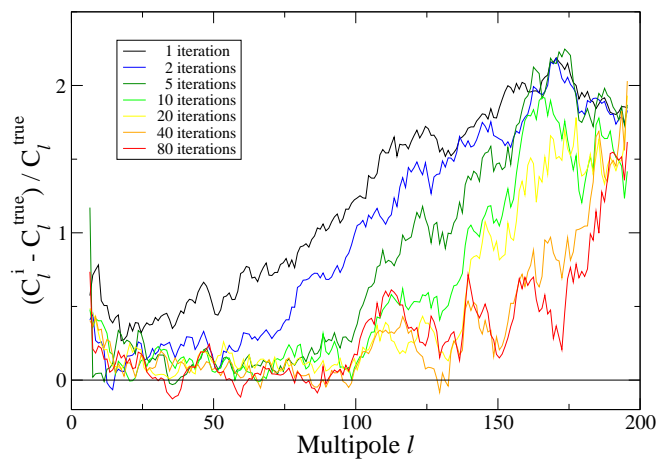}}

\caption{Burn-in time of the Gibbs sampler, computed from the
  low-resolution simulations. The initial guess was chosen to be three
  times the exact spectrum, and the sampler was run for 80
  iterations. Note how slowly the chain converges toward the correct
  region (i.e., toward the horizontal zero-axis) in the low
  signal-to-noise regime. A good initial guess is essential for the
  parallelization scheme proposed in this paper.}
\label{fig:burn_in}
\end{figure}

\begin{figure*}

\mbox{\epsfig{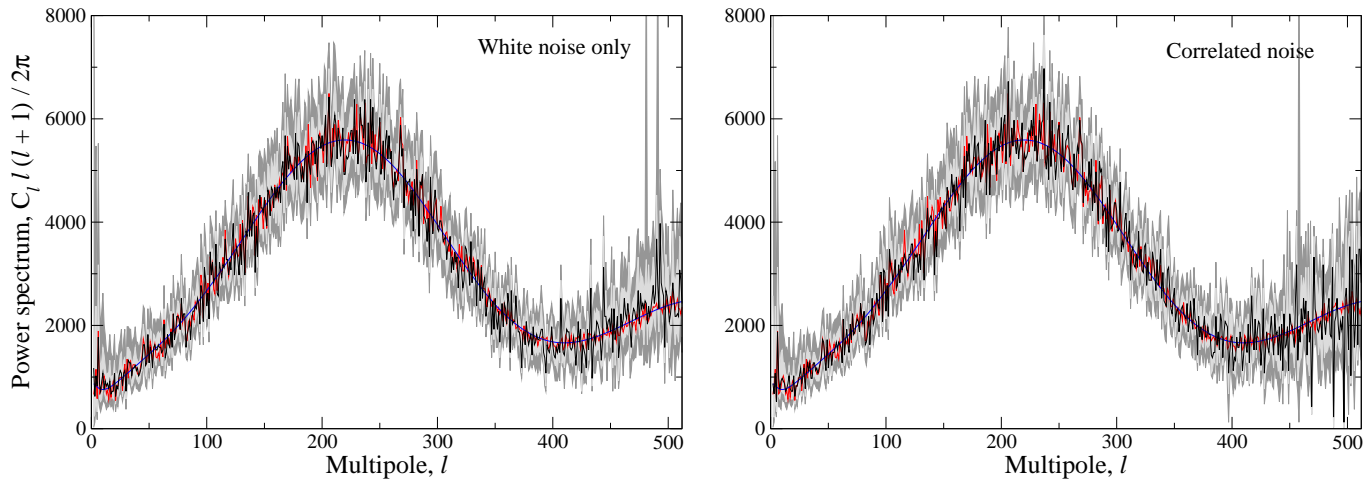}}

\caption{Results from high-resolution simulations. The red lines show
  the true power spectrum of the input realization, the blue lines
  show the ensemble averaged spectrum from which the CMB sky was
  drawn, and the black lines show the maximum likelihood spectra
  estimated by the Gibbs sampler. The gray bands indicate 1 and
  $2\sigma$ confidence regions. The spectra are computed from the
  combined Q+V+W simulations, taking into account individual beam and
  noise properties, and adding either white noise (left panel) or
  correlated noise (right panel) to the simulations.}
\label{fig:high_res_spec}
\end{figure*}

Finally, we turn to the issue of burn-in time. Although the theory of
Gibbs sampling guarantees us that the samples will converge toward
being samples from the joint distribution density, it does not tell us
when such convergence is obtained, and this must therefore be
established by experiments. We study this issue through a simple
exercise: Once again we utilize the simulated data described above,
but this time we choose a first power spectrum guess which is exactly
three times larger than the true spectrum. Then we run the algorithms
for a number of iterations, and plot the power spectrum samples as a
function of iteration count, $i$. The results from this exercise are
shown in figure \ref{fig:burn_in}, in the form of
\begin{equation}
x(i) = \frac{C_{\ell}^{i} - C_{\ell}^{\textrm{true}}}{C_{\ell}^{\textrm{true}}}.
\end{equation}
Note that the spectra have been averaged with a window width of
$\Delta \ell = 10$, making it easier to see the overall trends.

The conclusion to be drawn from this plot seems clear: A poor initial
guess can invalidate a large number of samples, and, in particular, a
weak estimate of the low signal-to-noise regime is very expensive to
correct. This can potentially pose a serious threat to our main
parallelization scheme, which is based on many independent short
chains, rather than one long chain. For this reason, the Gibbs
sampling approach in its current form may not be particularly well
suited as the only estimator for a new experiment. A faster method,
such as Master \citep{hivon:2002}, is therefore suggested to provide
an initial guess for the Gibbs samplers. Once an approximate power
spectrum is established, the Gibbs sampling process is already within
the appropriate range, and only a few samples need to be discarded, if
any at all. However, we do not need to rely blindly on the first
guess, since a poorly chosen starting point would lead to a systematic
drift in the Gibbs chains which should be easily detectable.

\subsection{High-resolution simulations}
\label{sec:highres_sim}

In this section we turn to high-resolution simulations, and undertake
a full-scale \emph{WMAP}-type analysis. The simulations in this case
are prepared in the same way as in the low-resolution case, except
with full-scale input data, and no inclusion of mono- and dipole
components.

The main limitation in this case is CPU time, and extremely long
chains are simply not feasible. Instead, we run many independent
chains in parallel, each producing only a small number of samples, as
discussed in section \ref{sec:parallelization}.

The analysis is designed to match the analysis of the first-year
\emph{WMAP} data presented by \citet{odwyer:2004}. Specifically, we
generate a random sky with the HEALPix utility synfast, and convolve
this sky with the beams corresponding to each of the eight \emph{WMAP}
bands (Q1--2, V1--2, W1--4). Next we add either white noise (with the
appropriate $N_{\textrm{obs}}$ patterns for each band) or correlated
noise (as generated by the \emph{WMAP} team\footnote{Available at
http://lambda.gsfc.nasa.gov.}) to these CMB maps. Finally, the
\emph{WMAP} Kp2 mask \citep{bennett:2003b} which excludes point
sources is imposed on the data, leaving 85\% of the sky available for
analysis. At this stage, the Gibbs sampler is run over 12
independently initialized chains for 60 iterations, for a total of 720
samples.

We point out that the numbers of observations per pixel,
$N_{\textrm{obs}}$, in the correlated noise files supplied by the
\emph{WMAP} team do not match perfectly those of the observed map
files, and unless the appropriate $N_{\textrm{obs}}$ patterns are used
in each case, a noise excess at $\ell \gtrsim 350$ is observed. The
white noise level, however, are identical for the two patterns, and so
this difference does not have a significant impact on the results, as
long as one is aware of the difference.

In figure \ref{fig:high_res_spec} we have plotted the power spectra
from the multiple-chain analysis, including white noise in the left
panel and correlated noise in the right panel. Overall, we see that
the agreement between the realization specific spectrum (red line) and
the maximum likelihood solution (black line) found by the Gibbs
sampler is quite good, and there is no detectable bias in any parts of
the spectrum.

\begin{figure}

\mbox{\epsfig{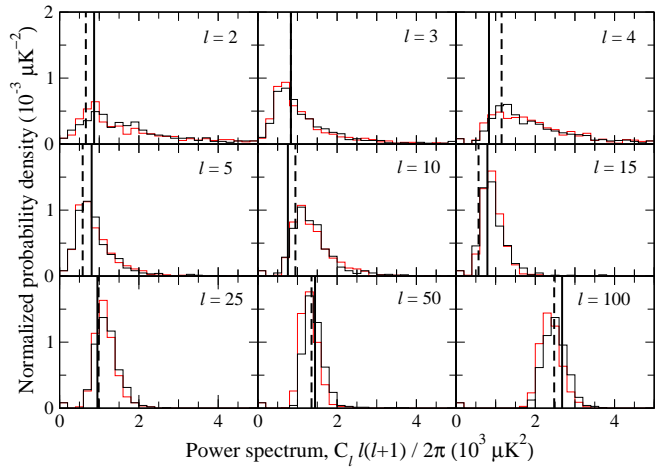}}

\caption{The $C_{\ell}$ distributions of a few selected multipoles,
  comparing the white noise (black histograms) to the correlated noise
  (red histograms) samples. The true realization specific spectrum is
  marked with a dashed vertical line, and the ensemble averaged
  spectrum with a solid line.}
\label{fig:wn_vs_cn_hist}
\end{figure}

Next, in figure \ref{fig:wn_vs_cn_hist} we plot a few selected
histograms of the power spectra, comparing the white (black
histograms) and correlated (red histograms) noise results more
directly. As we see, the agreement is generally very good, and in
particular, the three upper panels clearly demonstrate that the low
level of correlated noise present in the \emph{WMAP} data do not
compromise the low-$\ell$ spectrum.

At higher $\ell$'s, a small shift may be seen between the two
distributions, which is most likely due to the fact that the noise
realizations are different. We made similar plots for neighboring
multipoles, finding that the absolute levels of discrepancy seen in
figure \ref{fig:wn_vs_cn_hist} are quite typical for these angular
scales, and the signs of the shifts are random. Thus, the differences
does not seem to be indicative of a systematic bias.

By studying simulated data, we have thus explicitly demonstrated that
the Gibbs sampling technique is able to analyze the mega-pixel
\emph{WMAP} data set properly, and that neither correlated noise nor
incomplete sky coverage compromise the scientific results
significantly. All in all, the feasibility of this approach with
respect to current and future data sets has been firmly established.

\section{Conclusions}
\label{sec:conclusions}

We have implemented two independent versions of the Gibbs sampling
technique introduced by \citet{jewell:2004} and \citet{wandelt:2004},
and tested the performance and behavior of the codes thoroughly. In
particular, we have explicitly verified that the two implementations
produce identical output, despite a few algorithmic differences,
demonstrating that these algorithmic choices do not affect the
scientific results. Further, we applied the codes to simulated data
with controlled properties, and found the output to agree very well
with the theoretical expectations. In doing so, we also demonstrated
the feasibility of the method for high-resolution applications.

One of the main goals of these simulations was to build up intuition
about the phenomenological behavior of the Gibbs sampling algorithm,
focusing in particular on issues such as the correlation length of the
Markov chains, and the burn-in and convergence time. Through these
experiments, we found that the signal-to-noise ratio is by far the
most constraining factor to the algorithm in its current form. The
step size between two consecutive samples is determined by the cosmic
variance alone, while the overall posterior density incorporates noise
uncertainty as well. Thus, in the low signal-to-noise regime
subsequent samples are highly correlated, and the effective number of
independent samples is dramatically reduced. In its current
formulation, the method is therefore most efficient at scales for
which the signal-to-noise ratio is higher than, say, 0.5, or perhaps
up to $\ell = 350$--400 for the first-year \emph{WMAP} data. On the
other hand, the Gibbs sampler is only a special case from a more
general framework, and other sampling schemes may be introduced in
order to break these correlations. This will be the topic of a future
publication.

Perhaps the single most appealing feature of the Gibbs sampling
approach, is its ability to incorporate most real-world complications
in a statistically consistent manner. In this paper we have
demonstrated how to handle incomplete sky coverage and unknown mono-
and dipole contributions, which are the most important point for the
analysis of the first-year \emph{WMAP} data, but future extensions
will also include polarization, more sophisticated treatment of
foregrounds, internal sampling over cosmological parameters, inclusion
of asymmetric beams, and statistically consistent handling of $1/f$
noise. In fact, the Gibbs sampling approach is not simply a maximum
likelihood method, but rather a machinery facilitating an optimal,
global analysis. Needless to say, the computational challenges are
considerable, but with a scaling equivalent to that of map making
(which has to be performed in any approach currently proposed), this
method may just be able to do the job.

A second goal of this paper was to prepare for the actual analysis of
the \emph{WMAP} data, by applying the algorithm to simulated data with
similar properties. Specifically, we showed that the estimated power
spectrum is unbiased, and that even the lowest-order multipoles are
not compromised by the either the galactic cut, given that the
foreground correction method presented by \citet{bennett:2003a} is
adequate, or by the (low levels of) correlated noise present in the
data. Thus, no further sophistications beyond those presented in this
paper seem necessary in order to perform a valid Bayesian analysis of
the first-year \emph{WMAP}. The scientific results from this analysis
are presented in a companion letter by \citet{odwyer:2004}.

\begin{acknowledgements}
We acknowledge use of the HEALPix software (G\'orski, Hivon \& Wandelt
1998) and analysis package for deriving the results in this paper. We
also acknowledge use of the Legacy Archive for Microwave Background
Data Analysis (LAMBDA). H.\ K.\ E.\ and P.\ B.\ L.\ acknowledge
financial support from the Research Council of Norway, including a
Ph.\ D.\ studentship for H.\ K.\ E. This work has also received
support from The Research Council of Norway (Programme for
Supercomputing) through a grant of computing time. This work was
partially performed at the Jet Propulsion Laboratory, California
Institute of Technology, under a contract with the National
Aeronautics and Space Administration. This work was partially
supported by an NCSA Faculty Fellowship for B.\ D.\ W. This research
used resources of the National Energy Research Scientific Computing
Center, which is supported by the Office of Science of the
U.S. Department of Energy under Contract No. DE-AC03-76SF00098.
\end{acknowledgements}

\end{document}